\documentclass[a4paper,12pt]{article}
\pdfoutput=1 

\usepackage{amsmath,amssymb}
\usepackage[pdftex]{hyperref}
\hypersetup{colorlinks=true,linkcolor=blue,urlcolor=blue,filecolor=blue,citecolor=blue}
\usepackage{epsf}
\usepackage{cite}
\usepackage{fancyhdr}
\usepackage{enumerate} 
\usepackage{bm}
\usepackage[pdftex]{graphicx}

\usepackage[eulergreek]{sansmath}
\def\sfgamma{\mbox{\sffamily\sansmath$\mathrm{\gamma}$}}
\def\sfTheta{\mbox{\sffamily\sansmath$\mathrm{\Theta}$}}
\def\sfSigma{\mbox{\sffamily\sansmath$\mathrm{\Sigma}$}}
\def\sfa{\mathsf{a}}
\def\sfk{\mathsf{k}}
\def\sfA{\mathsf{A}}
\def\sfB{\mathsf{B}}
\def\sfZ{\mathsf{Z}}

\usepackage[usenames,dvipsnames]{xcolor}

\makeatletter

\g@addto@macro\bfseries{\boldmath}

\def\refchecklabelfontsize{\fontsize{5pt}{5pt}\selectfont}
\let\mark@size=\refchecklabelfontsize

\def\half{{\frac{1}{2}}}
\def\thalf{{\tfrac{1}{2}}}
\def\p{\partial}

\def\unit{{1\kern-.65ex {\rm l}}}
\def\1{{1\kern-.65ex {\rm l}}}
\def\ap{{\alpha'}}

\def\ket#1{{|{#1}\rangle}}

\def\kett#1{{|{#1}\rangle\!\rangle}}

\let\ev=\bracket

\def\bh{{\widehat{b}}}

\def\Ah{{\widehat{A}}}
\def\Bh{{\widehat{B}}}

\def\Nh{{\widehat{N}}}

\def\muh{{\widehat{\mu}}}
\def\varthetah{{\widehat{\vartheta}}}

\def\htilde{{\widetilde{h}}} 
\def\jt{{\widetilde{j}}}

\def\Gt{{\widetilde{G}}}

\def\Jt{{\widetilde{J}}}
\def\Lt{{\widetilde{L}}}

\def\phit{{\widetilde{\phi}}}
\def\psit{{\widetilde{\psi}}}

\def\zb{{\overline{z}}}

\def\Nb{{\overline{N}}}

\def\cA{{\cal A}}
\def\cB{{\cal B}}
\def\cC{{\cal C}}
\def\cD{{\cal D}}

\def\cF{{\cal F}}

\def\cJ{{\cal J}}
\def\cK{{\cal K}}
\def\cL{{\cal L}}
\def\cM{{\cal M}}
\def\cN{{\cal N}}
\def\cO{{\cal O}}
\def\cP{{\cal P}}

\def\cR{{\cal R}}

\def\bbR{{\mathbb{R}}}

\def\bbZ{{\mathbb{Z}}}

\newcount\hour \newcount\minute
\hour=\time \divide \hour by 60
\minute=\time
\def\now{%
\ifnum \hour<13
  \ifnum \hour=0 \advance \hour by 12 \number\hour:\else \number\hour:\fi%
     \ifnum \minute<10 0\fi%
     \number\minute%
\ A.M.%
\else \advance \hour by -12 \number\hour:%
  \ifnum \minute<10 0\fi%
  \number\minute%
  \ P.M.%
\fi%
}

\makeatother

\begin{document}

\baselineskip=18pt  
\numberwithin{equation}{section}  

%


%



\thispagestyle{empty}

\vspace*{-2cm} 
\begin{flushright}
YITP-20-16\\
\end{flushright}

\vspace*{2.5cm} 
\begin{center}
 {\LARGE Superstrata}\\
 \vspace*{1.7cm}
 Masaki Shigemori\\
 \vspace*{1.0cm} 
Department of Physics, Nagoya University, Nagoya 464-8602\\
and\\
Yukawa Institute for Theoretical Physics, Kyoto University\\
Kitashirakawa Oiwakecho, Sakyo-ku, Kyoto 606-8502 Japan
\end{center}
\vspace*{1.5cm}

\noindent

We give a survey of the present status of the microstate geometries
called superstrata.  Superstrata are smooth, horizonless solutions of
six-dimensional supergravity that represent some of the microstates of
the D1-D5-P black hole in string theory.  They are the most general
microstate geometries of that sort whose CFT dual states are identified.
After reviewing relevant features of the dual CFT, we discuss the
construction of superstratum solutions in supergravity, based on the
linear structure of the BPS equations.  We also review some of recent
work on generalizations of superstrata and physical properties of
superstrata.  Although the number of superstrata constructed so far is
not enough to account for the black-hole entropy, they give us valuable
insights into the microscopic physics of black holes.

\newpage
\setcounter{page}{1} 



\tableofcontents


\newpage
\section{Introduction}

The microstate geometry program aims to explicitly construct as many
microstates of black holes as possible, as ``microstate geometries'',
{\it i.e.\/}\ smooth, horizonless solutions of classical supergravity.
In this program, the so-called D1-D5 system has played an important
role.  This system is obtained by compactifying type IIB string theory
on~$S^1\times \cM $ with $\cM =T^4$ or K3 and wrapping $N_1$
D1-branes\footnote{For $\cM={\rm K3}$, $N_1$ includes the D1-brane
charge induced on the worldvolume of the D5-branes by a curvature
coupling \cite{Bershadsky:1995qy}.  Namely $N_1=N_1^{\rm
explicit}+N_1^{\rm induced}$, $N_1^{\rm induced}=-N_5$.  } on~$S^1$ and
$N_5$ D5-branes on~$S^1\times \cM $.  The size of $\cM$ is taken to be
of the string scale while the radius $R_y$ of $S^1$ remains macroscopic.  If
we add a third charge,~$N_P$ units of Kaluza-Klein momentum (P) charge
along~$S^1$, we have a 1/8-BPS, 3-charge black hole in five dimensions
(or black string in six dimensions, if we include~$S^1$) with a finite
entropy which was reproduced by Strominger and Vafa by counting
microstates in the brane worldvolume theory \cite{Strominger:1996sh}.
More generally, we can also add left-moving angular momentum $J$ and the
area entropy of the resulting 1/8-BPS black hole (the BMPV black hole
\cite{Breckenridge:1996is}) is given by\footnote{$J=J^3_0\in\bbZ/2$ where
$J^3_0$ is a generator of $SU(2)_L\in SO(4)$ coming from the rotational
symmetry in the directions transverse to the D-branes.}
\begin{align}
 S_{\rm BMPV}=2\pi\sqrt{N_1 N_5 N_P-J^2}.\label{S_BMPV}
\end{align}
A central question in the microstate geometry program is how much of
this entropy can be accounted for by supergravity solutions.

The precursor of the microstate geometry program was the study of the
2-charge states of the D1-D5 system, namely the ones with $N_P=0$.  In
this case, the microstates can be realized as microstate geometries
called Lunin-Mathur geometries \cite{Lunin:2001jy, Lunin:2002iz,
Taylor:2005db, Kanitscheider:2007wq}, which are parametrized by
functions of one variable.  The growth of the microscopic entropy,
$S_{\text{2-chg}}\sim \sqrt{N_1 N_5}$, can be reproduced by counting
Lunin-Mathur geometries \cite{Rychkov:2005ji, Krishnan:2015vha},
although the 2-charge ensemble has vanishing area entropy at the
classical level.

This success led to the microstate geometry program to construct
microstate geometries for the 3-charge system, namely for the case with
$N_P>0$, for which the area entropy is non-vanishing at the classical
level.  Many families of 3-charge microstates have been constructed
based on five-dimensional multi-center solutions \cite{Bena:2006kb,
Bena:2010gg, Heidmann:2017cxt, Bena:2017fvm} (see \cite{Bena:2005va,
Berglund:2005vb} about smooth multi-center solutions) and other methods,
such as solution-generating technique, the matching technique, and BPS
equations \cite{Mathur:2003hj, Lunin:2004uu, Giusto:2004id,
Giusto:2004ip, Giusto:2006zi, Ford:2006yb, Mathur:2011gz, Mathur:2012tj,
Lunin:2012gp, Giusto:2013bda, Giusto:2012yz} (see also
\cite{Bena:2013pda, Warner:2019jll}).  More recently, a new
class of microstate geometries called \emph{superstrata} was constructed
\cite{Bena:2015bea}.  Superstrata are solutions of six-dimensional
supergravity parametrized by functions of three variables, and represent
the most general microstate geometries known thus far for the D1-D5-P
black hole, with understood CFT dual.\footnote{The CFT dual states of
superstrata based on the five-dimensional multi-center solutions with
two centers are known, while duals of multi-center solutions with more
than two centers or those of their superstratum generalizations are not
known.}

The existence of superstrata was conjectured based on the idea of double
supertube transition \cite{Bena:2011uw}. The 2-charge microstates
(Lunin-Mathur geometries) have a dipole charge which does not exist in
the original configuration of the D1- and D5-branes but is produced by
the supertube transition \cite{Mateos:2001qs}. In \cite{deBoer:2010ud,
deBoer:2012ma}, it was argued that, via a multistage supertube
transition, black-hole microstates involve various dipole charges that
do not exist in the original configuration.\footnote{The string-theory
configurations resulting from such multistage supertube transition 
should be called (general) superstrata, which in general contain dipole
charges that do not allow a description in terms of smooth geometry.
Superstrata that do allow a geometric description, like the ones
constructed in \cite{Bena:2015bea}, should properly be called
\emph{geometric} superstrata, although they are normally simply called
superstrata.}  Based on this idea, in \cite{Bena:2011uw}, it was
conjectured that the double supertube transition in the D1-D5-P system will
lead to smooth microstate geometries, {\it i.e.\/}\ superstrata, that
depend on the coordinate of the $S^1$, $v$, and are parametrized by
functions of (at least) two variables.  This $v$-dependence means that
superstrata will live in six-dimensional supergravity.

Although it was conjectured that superstrata exist in six-dimensional
supergravity, there were some more steps needed for their actual
construction. The first was the realization that the BPS equations of
six-dimensional supergravity \cite{Gutowski:2003rg, Cariglia:2004kk}
have a linear structure \cite{Bena:2011dd}; namely, the BPS equations
can be organized so that they are linear if solved in a certain order.
Leveraging this structure, one can start with infinitesimal (linear)
perturbation around a simple background and then non-linearly complete
it to obtain a fully backreacted solution.
Another step is so-called ``coiffuring'' \cite{Bena:2015bea}. This means
that, when one goes from a linear (infinitesimal) solution to a full
non-linear solution, one must turn on fields that were not turned on in
the linear solution, in order for the geometry to be regular (plus, in
order for the BPS equations to simplify and be solvable). From an
AdS/CFT viewpoint, this is due to the fact that \cite{Giusto:2019qig}
single-trace operators and double-trace operators with the same quantum
numbers mix and therefore turning on one at linear order implies turning
on the other at higher order. The fields particularly relevant for
coiffuring were identified in \cite{Giusto:2011fy, Giusto:2012jx,
Giusto:2013rxa}.

Based on these developments, the first examples of superstrata were
explicitly constructed in \cite{Bena:2015bea}, providing a proof of
their existence. After that, superstratum solutions have been
generalized in many ways \cite{ Bena:2016agb, Bena:2016ypk,
Bena:2017geu, Bena:2017xbt, Bakhshaei:2018vux, Ceplak:2018pws,
Heidmann:2019zws}, and a number of checks of their proposed AdS/CFT
dictionary were carried out \cite{Giusto:2015dfa, Giusto:2019qig}.  By
now it is fair to say that we have a pretty good picture of possible
superstrata solutions, although explicit expressions have been found
only for a limited class of solutions.  They come in multiple species
and are parametrized by functions of three variables, with rich physical
content that one can explore.  Various physical aspects of the solutions
and their implications for black-hole microphysics are being actively
investigated; instead of listing the relevant literature here, we will
review some of the developments later in this article.  It is expected
that superstrata will continue to give us useful insights into the
microstructure of black holes in string theory.

In the remainder of the article, we will review the construction of
superstrata with some explicit examples and then attempt a short survey
of further developments in the literature.
In section~\ref{sec:CFT}, we review the CFT picture of the states of the
D1-D5 system, focusing on the map between chiral primary states in CFT
and 1/4-BPS supergraviton states in the bulk, and between left-descendant
states in CFT and 1/8-BPS supergraviton states in the bulk.  Superstrata
are nothing but coherent superpositions of 1/8-BPS supergravitons.
In section~\ref{sec:sugra_setup}, we review the supergravity setup and
the three layers of equations to be satisfied by supersymmetric
solutions. We discuss some 1/4-BPS microstate geometries as examples.
In section~\ref{sec:superstrata}, we review the construction of
superstrata based on the formulation of section~\ref{sec:sugra_setup}.
We emphasize the importance of focusing on solutions for which the base
space is fixed and, taking flat base, construct explicit solutions.
Superstrata can have arbitrary sets of modes but, for simplicity, we
focus on the single-mode superstrata in which only one mode is turned
on.
In section~\ref{sec:developments}, we review some of the recent developments
in generalizing superstratum solutions and the studies of their properties.
We end with a conclusion in section \ref{sec:conclusion}.

\section{CFT}
\label{sec:CFT}

After the decoupling limit, the geometry of the D1-D5 system becomes
asymptotically AdS$_3\times S^3\times \cM$.  This means that this system
can be equivalently described by a holographic CFT, which is known as
the D1-D5 CFT\@.\footnote{For reviews of the D1-D5 CFT, see {\it
e.g.\/}~\cite{David:2002wn, Avery:2010qw}.}  This theory is a
$d=2,\cN=(4,4)$ CFT with a symmetry group $SU(1,1|2)_L\times
SU(1,1|2)_R$, which is generated by the affine generators $L_n,G^{\alpha
A}_n,J^i_n$ and their right-moving versions $\Lt_n,\Gt^{\dot{\alpha}
A}_n,\Jt^{\,\bar{i}}_n$.  Here, $\alpha=\pm$ is a doublet index and
$i=1,2,3$ is a triplet index for $SU(2)_L\subset SU(1,1|2)_L$, while
$\dot{\alpha},\bar{i}$ are their right-moving counterparts.  The index
$A=1,2$ is the doublet index for an additional $SU(2)_B$ symmetry group
which acts as an outer automorphism on the superalgebra.  In its moduli
space, the D1-D5 CFT is believed to have an orbifold point where the
theory is described by a supersymmetric sigma model with the target
space being the symmetric orbifold, ${\rm Sym}^N \cM$, where
\cite{Vafa:1995bm, Witten:1997yu}\footnote{For $\cM=T^4$, the low-energy
dynamics of the D-brane bound state can be described by a supersymmetric
sigma model with target space $\bbR^4\times T^4\times {\rm Sym}^{N_1
N_5}(T^4)$, where the $\bbR^4$ part describes the center-of-mass motion
of the D-branes in the noncompact $\bbR^4$, the $T^4$ part describes
worldvolume Wilson lines along the internal $T^4$, and the ${\rm
Sym}^{N_1 N_5}(T^4)$ part describes the moduli space of D1-branes as
instantons inside the D5 worldvolume \cite{Vafa:1995bm,
Maldacena:1999bp}.  Here we are focusing on the last part.  For
$\cM={\rm K3}$, the target space is $\bbR^4\times {\rm Sym}^{N_1
N_5+1}({\rm K3})$ \cite{Vafa:1995bm, Strominger:1996sh} where the
$\bbR^4$ part describes the center-of-mass motion in the noncompact
$\bbR^4$ and the ${\rm Sym}^{N_1 N_5+1}({\rm K3})$ part describes
the instanton moduli space which we are focusing on.}
\begin{align}
 N\equiv 
\begin{cases}
 N_1 N_5  & (\cM=T^4),\\
 N_1 N_5+1  & (\cM={\rm K3}).\\
\end{cases}
\end{align}
We will be
working at the orbifold point henceforth.

If we want to preserve supersymmetry in the D1-D5 system, we
must impose the periodic boundary condition for the worldvolume fermions
along the $S^1$ on which the D-branes are wrapped.  This means that we
are naturally in the RR (Ramond-Ramond) sector of the SCFT\@.  

In this section, we review the structure of the BPS states in the D1-D5
CFT, whose holographic dual we are after.  We start with the NS
(Neveu-Schwarz) sector of the theory in which the spectrum of states is
somewhat more transparent, and then discuss the R sector.

\subsection{NS sector}

\subsubsection{1/4-BPS supergraviton states}

In the NS-NS sector, the theory has the unique vacuum with
$L_0=\tilde{L}_0=0$ which preserves all the $8+8$ supercharges of the
theory. The bulk dual of the NS-NS vacuum is empty ${\rm AdS}_3\times
S^3$.

As excited states above the NS-NS vacuum, the theory has
\emph{single-particle} chiral primary states which are in
one-to-one correspondence with the Dolbeault cohomology of $\cM $
\cite{Maldacena:1998bw, Vafa:1994tf}.  For $\cM =T^4$, we have 16
species of states
\begin{align}
T^4:\qquad
\begin{aligned}
 &\ket{\alpha\dot\alpha}_k,&
 h&=j=\tfrac{k+\alpha}{2},& 
 \htilde&=\jt=\tfrac{k+\dot{\alpha}}{2},&&\text{bosonic},\\
 &\ket{\alpha \dot{A}}_k,&
 h&=j=\tfrac{k+\alpha}{ 2},&
 \htilde&=\jt=\tfrac{k}{ 2},&&\text{fermionic},\\
 &\ket{\dot{A}\dot\alpha}_k,&
 h&=j=\tfrac{k}{ 2},&
 \htilde&=\jt=\tfrac{k+\dot\alpha}{ 2},&&\text{fermionic},\\
 &\ket{\dot{A}\dot{B}}_k,&
 h&=j=\tfrac{k}{ 2},& 
 \htilde&=\jt=\tfrac{k}{ 2},&&\text{bosonic},
\end{aligned}
\label{T4_1-part_ch_pr}
\end{align}
where $k=1,\dots,N$.  $\dot{A},\dot{B}=1,2$ are doublet indices for an
$SU(2)_C$ that is not part of the symmetry group of the theory. $h,j$
are the values of $L_0,J_0^3$, while $\htilde,\jt$ are those of
$\Lt_0,\Jt_0^3$.  At the orbifold point, these states correspond to
twist operators of order $k$; namely, they intertwine $k$ copies of $\cM
$ (out of $N$ copies).  We refer to these $k$ copies, thus intertwined
together, as a strand of length $k$.  Because spin is $j-\jt$, the
states $\ket{\alpha\dot\alpha},\ket{\dot{A}\dot{B}}$ are bosonic while
$\ket{\alpha\dot A},\ket{\dot{A}\dot{\alpha}}$ are fermionic.  The
$SU(2)_{C\,}$-invariant linear combination $
\tfrac{1}{\sqrt{2}}\epsilon_{\dot{A}\dot{B}}\ket{\dot{A}\dot{B}}_k$ is
denoted by $\ket{00}_k$, which corresponds to the K\"ahler form of $T^4$.
For $\cM={\rm K3}$, there are 24 species of single-particle chiral primary states and
they are all bosonic:
\begin{align}
{\rm K3}:\qquad
\begin{aligned}
 &\ket{\alpha\dot\alpha}_k,&
 j&=\tfrac{k+\alpha}{2},& 
 \jt&=\tfrac{k+\dot{\alpha}}{2}\\
 &\ket{I}_k,&
 j&=\tfrac{k}{2},& 
 \jt&=\tfrac{k}{2},\qquad I=1,\dots,20.\\
\end{aligned}
\label{K3_1-part_ch_pr}
\end{align}
Among $\ket{I}_k$, the one that corresponds to the K\"ahler form of K3
is 
denoted by $\ket{00}_k$.

All these states \eqref{T4_1-part_ch_pr}, \eqref{K3_1-part_ch_pr}
preserve 8 supercharges, 4 from the left and another 4 from the
right.\footnote{Except for the case with $\alpha=-$ ($\dot{\alpha}=-$)
and $k=1$ for which 8 left-moving (right-moving) supercharges are
preserved.}  Conventionally, they are said to be 1/4-BPS, relative to
the amount of supersymmetry (32 supercharges) of type IIB superstring in
ten dimensions (although they preserve half of the supersymmetry of the
D1-D5 CFT)\@.

Among the states in \eqref{T4_1-part_ch_pr}, \eqref{K3_1-part_ch_pr},
the state $\ket{--}_1=\ket{\alpha=-,\dot\alpha=-}_1$ is special because
it has $h=j=\htilde=\jt=0$ and actually represents the vacuum (of a
single copy of $\cM $).  All other states can be thought of as
excitations and, via AdS/CFT, correspond to the possible excitations in
linearized supergravity around empty AdS$_3\times S^3$, called
``supergravitons''. In other words, each of the chiral primary states
\eqref{T4_1-part_ch_pr}, \eqref{K3_1-part_ch_pr} (except $\ket{--}_1$)
is in one-to-one correspondence with a particular single-particle, 1/4-BPS
state of the supergraviton propagating in the bulk AdS$_3\times S^3$
background \cite{Deger:1998nm, Maldacena:1998bw, Larsen:1998xm,
deBoer:1998kjm}.

If we multiply together single-particle chiral primary states, we obtain
\emph{multi-particle} chiral primary states, which are the most general
1/4-BPS states.  Explicitly, they can be written as
\begin{align}
 \prod_{\psi}
 \prod_{k=1}^N
 \bigl[\ket{\psi}_k\bigr]^{N^{\psi}_k},
\label{gen_ch_pr_cft}
\end{align}
where $\ket{\psi}$ runs over different species in
\eqref{T4_1-part_ch_pr} or \eqref{K3_1-part_ch_pr}.  The general chiral
primary state is specified by the set of numbers $\{N^\psi_k\}$, which
correspond to the number of strands of species~$\ket{\psi}$ and length
$k$.  The values that $N^\psi_k$ can take are $0,1,2,\dots$ if
$\ket{\psi}$ and $0,1$ if $\ket{\psi}$ is fermionic. The strand numbers
$\{N^\psi_k\}$ must satisfy the constraint that the total strand length is
equal to $N$:
\begin{align}
 \sum_\psi \sum_k k N^\psi_{k}=N.
\label{sum_kN_k=N}
\end{align}
The non-trivial part of the multi-particle chiral primary state
\eqref{gen_ch_pr_cft} is made of single-particle chiral primary states that
are not the trivial state $\ket{--}_1$.  The trivial part of the state
is made of $N^{--}_1$ copies of the trivial state $\ket{--}_1$, so that the total strand length is $N$.

In the bulk, the states \eqref{gen_ch_pr_cft} correspond to
multi-particle, 1/4-BPS states of supergravitons (``supergraviton
gas'').  Namely, the states \eqref{gen_ch_pr_cft} span the Fock space of
1/4-BPS supergravitons, modulo the constraint \eqref{sum_kN_k=N}.  When
$N^{\psi}_k=\cO(N)$ (where $\ket{\psi}_k\neq \ket{--}_1$), the bulk
picture of supergravitons propagating in undeformed AdS$_3\times S^3$
is no longer valid but the geometry becomes deformed by backreaction.

In order to correspond to the supergravity point, the boundary CFT must
be perturbed away from the orbifold point where the chiral primary states
have the above simple description.  Even if we go away from the orbifold
point, the number of chiral primary states remains the same, although
individual states can mix into each other
\cite{deBoer:2008qe}.\footnote{For $\cM={\rm K3}$, supersymmetry implies
that the number of chiral primary states do not change
\cite{Aharony:1999ti}. For $\cM=T^4$, such supersymmetry argument is not
enough for showing that the number stays constant, although we expect
that it does, on physical grounds (single-particle supergravitons and
their gas must exist everywhere in the moduli space).  }

\subsubsection{1/8-BPS supergraviton states}

The single-particle chiral primary states in \eqref{T4_1-part_ch_pr} and
\eqref{K3_1-part_ch_pr} are the highest-weight states with respect to
the rigid $SU(1,1|2)_L\times SU(1,1|2)_R$ symmetry and more general,
descendant states in the $SU(1,1|2)_L\times SU(1,1|2)_R$ multiplet can
be obtained by the action of the rigid generators
$\{L_{-1},G_{-1/2}^{-,A},J_0^-\}$ and
$\{\Lt_{-1},\Gt_{-1/2}^{-,A},\Jt_0^-\}$.  To preserve supersymmetry, we
will only consider descendants obtained by the action of the left-moving
generators $\{L_{-1},G_{-1/2}^{-,A},J_0^-\}$.  If we start with a chiral
primary states with $h=j$, which we denote by $\kett{j,j}$, we generate the
following states:
\begin{align}
\begin{split}
 & \textstyle
\kett{j+n,j}\xrightarrow{J_0^-}\kett{j+n,j-1}\xrightarrow{J_0^-}\cdots\xrightarrow{J_0^-}\kett{j+n,-j}
 \\
 &G_{-1/2}^{-A}\bigg\downarrow\\
 &\textstyle
 \kett{j+\half+n,j-\half}\xrightarrow{J_0^-}\kett{j+\half+n,j-{3\over 2}}\xrightarrow{J_0^-}\cdots\xrightarrow{J_0^-}\kett{j+\half+n,-(j-\half)}
 \\
 &G_{-1/2}^{-B}\bigg\downarrow\\
 &\textstyle
 \kett{j+1+n,j-1}\xrightarrow{J_0^-}\kett{j+1+n,j-2}\xrightarrow{J_0^-}\cdots\xrightarrow{J_0^-}\kett{j+1+n,-(j-1)}
\end{split}
\label{members_of_short_multiplet}
\end{align}
Here, $\kett{h,j}$ means a state with $(L_0,J_0^3)=(h,j)$.  The states
in the second line are doubly degenerate, because we can use
$G_{-1/2}^{-A}$ with either $A=1$ or $A=2$ to descend from the first
line to the second.  The third line has no such degeneracy because we
can only descend from the first line with
$G_{-1/2}^{-,1}G_{-1/2}^{-,2}$.  More precisely, to get a genuinely new
state, we must act instead with $G_{-1/2}^{-,1}G_{-1/2}^{-,2}+{1\over
2h}L_{-1}J_0^-$ where $h$ is the value of $L_0$ for the chiral primary state
\cite{Avery:2010qw,Ceplak:2018pws}.  Moreover, the number $n=0,1,\dots$
corresponds to the number of times we act on the state with $L_{-1}$.
We denote the states thus obtained building on $\ket{\psi}_k$
by\footnote{These states are not normalized.}
\begin{subequations} 
\label{1p_1/8_sugrtn}
  \begin{align}
 \ket{\psi;k,m,n}   &=\tfrac{1}{m!\,n!}(J_0^-)^m (L_{-1})^n \ket{\psi}_k,\label{1p_1/8_sugrtn1}\\[1ex]
 \ket{\psi;k,m,n,A} &=\tfrac{1}{(m-1/2)!\,(n-1/2)!}(J_0^-)^{m-1/2} (L_{-1})^{n-1/2} G_{-1/2}^{-,A} \ket{\psi}_k,\label{1p_1/8_sugrtn2}\\[1ex]
 \ket{\psi;k,m,n,12}&=\tfrac{1}{(m-1)!\,(n-1)!}(J_0^-)^{m-1} (L_{-1})^{n-1} \left(G_{-1/2}^{-,1}G_{-1/2}^{-,2}+\tfrac{1}{2h}L_{-1}J_0^-\right) \ket{\psi}_k.\label{1p_1/8_sugrtn3}
 \end{align}
\end{subequations}
The range of $m,n$ is: $m=0,1,\dots,2h$, $n=0,1,2,\dots$ for
\eqref{1p_1/8_sugrtn1}; $m={1\over 2},{3\over 2},\dots,2h-{1\over 2}$,
$n={1\over 2}, {3\over 2}, \dots$ for \eqref{1p_1/8_sugrtn2}; and
$m=1,2,\dots,2h-1$, $n=1,2,3,\dots$ for \eqref{1p_1/8_sugrtn3}.  The
numbers $m$ and $n$ give the increase in $-J_0^3$ and $L_0$ relative to
the chiral primary state $\ket{\psi}_k$.  If $h=0$ the states
\eqref{1p_1/8_sugrtn2} and \eqref{1p_1/8_sugrtn3} do not exist, and if
$h=1/2$ the state \eqref{1p_1/8_sugrtn3} does not exist.  If the chiral
primary state $\ket{\psi}_k$ is bosonic (fermionic), the states
\eqref{1p_1/8_sugrtn1} and \eqref{1p_1/8_sugrtn3} are bosonic
(fermionic) while the state \eqref{1p_1/8_sugrtn2} is fermionic
(bosonic).  These states break all left-moving supersymmetry but
preserve 4 right-moving supercharges.  In the bulk, they correspond to
single-particle, 1/8-BPS supergraviton states obtained by the bulk
action of the rigid $SU(1,1|2)_L$ generators.  
Although we only considered descendants obtained by left rigid
generators here, if we also included descendants obtained by right rigid
generators, we could reproduce the complete spectrum of linearized
supergravity around AdS$_3\times S^3$ \cite{Deger:1998nm,
Maldacena:1998bw, Larsen:1998xm, deBoer:1998kjm}.

Just as in the 1/4-BPS case, we can multiply together single-particle
1/8-BPS states to construct a more general, multi-particle 1/8-BPS state:
\begin{align}
 \prod_{\psi,k,m,n,f} \bigl[\ket{\psi;k,m,n,f}\bigr]^{N^{\psi}_{k,m,n,f}},\qquad
 \sum_{\psi,k,m,n,f} k N^{\psi}_{k,m,n,f}=N,\qquad
 \label{gen_1/8_sugrtn_cft}
\end{align}
where $f={\rm null},A,12$ so that it covers all the three kinds in
\eqref{1p_1/8_sugrtn}. If the state $\ket{\psi;k,m,n,f}$ is bosonic
(fermionic), $N^\psi_{k,m,n,f}=0,1,2,\dots$ ($N^\psi_{k,m,n,f}=0,1$).
The state \eqref{gen_1/8_sugrtn_cft} corresponds in the bulk to a
1/8-BPS state of the supergraviton gas.  Namely,
\eqref{gen_1/8_sugrtn_cft} spans the Fock space of 1/8-BPS
supergravitons, modulo the constraint on $N^\psi_{k,m,n,f}$.

\subsection{R sector}

By spectral flow transformation, we can map all the above statements
into the R sector, which more directly corresponds to the bulk states of
the D1-D5 system. By spectral transformation, the charges $(h,j)$ of a
state on a strand of length $k$ are transformed as follows:
\begin{align}
h'&=h+2\eta j+k\eta^2,\qquad 
j'=j+{k}\eta.\label{spfl}
\end{align}
If we take the flow parameter $\eta=-1/2$, NS states get mapped into R states.
However, to match the convention of charges to that in the literature
\cite{Bena:2015bea, Bena:2016ypk, Bena:2017xbt}, we further flip the
sign of the $SU(2)_L$ charge, as $j\to -j$. So, the map from NS to R
that we will be using is
\begin{align}
h^{\rm R}&=h^{\rm NS}-j^{\rm NS}+{k\over 4},\qquad 
j^{\rm R}={k\over 2}-j^{\rm NS}.
\label{spfl+flip}
\end{align}
The same transformation in the right-moving sector is understood.

The map \eqref{spfl+flip} transforms single-particle chiral primary states into
R ground states on a single strand of length $k$.  For example,
\begin{align}
\label{NS-R_map_example}
\begin{aligned}
 &\ket{--}_k^{\rm NS},&& h^{\rm NS}=j^{\rm NS}=\tfrac{k-1}{2}
 &&\to \quad
 \ket{++}_k^{\rm R},&&
 h^{\rm R}=\tfrac{k}{4},
 j^{\rm R}=\tfrac{1}{2},
 \\
 &\ket{00}_k^{\rm NS},&& h^{\rm NS}=j^{\rm NS}=\tfrac{k}{2}
 &&\to \quad
 \ket{00}_k^{\rm R},&&
 h^{\rm R}=\tfrac{k}{4},
 j^{\rm R}=0.
\end{aligned}
\end{align}
The NS vacuum (empty ${\rm AdS}_3\times S^3$ in the bulk) goes to the
following R ground state:
\begin{align}
[\ket{--}_1^{\rm NS}]^N  \quad\to \quad
[\ket{++}_1^{\rm R}]^N
\label{NSvac-R}
\end{align}
The general R ground states, which are general 1/4-BPS states, are
\begin{align}
 \prod_{\psi}
 \prod_{k=1}^N
 \left[\ket{\psi}_k^{\rm R}\right]^{N^{\psi}_k},\qquad
 \sum_\psi \sum_k k N^\psi_{k}=N.
\label{1/4_sgrvtn_R}
\end{align}
where $\ket{\psi}$ runs over the species in \eqref{T4_1-part_ch_pr} or
\eqref{K3_1-part_ch_pr}, now understood as R ground states on a strand
of length $k$.  Coherent superpositions \cite{Skenderis:2006ah,
Kanitscheider:2006zf, Kanitscheider:2007wq} of these supergraviton
states are dual to smooth 1/4-BPS geometries called Lunin-Mathur
geometries \cite{Lunin:2001jy, Lunin:2002iz, Taylor:2005db,
Kanitscheider:2007wq}, as mentioned in the introduction.

The 1/8-BPS states in the NS sector, \eqref{gen_1/8_sugrtn_cft}, map
into the R states of the following form:
\begin{subequations} 
 \label{1/8_sgrvtn_R}
 \begin{gather}
 \Psi(\{N^\psi_{k,m,n,f}\})\equiv 
 \prod_{\psi,k,m,n,f} \Bigl[\ket{\psi;k,m,n,f}^{\rm R}\Bigr]^{N^{\psi}_{k,m,n,f}},\\
 \sum_{\psi,k,m,n,f} k N^\psi _{k,m,n,f}=N,
 ,\label{1/8_sgrvtn_R_constraint}
 \end{gather}
\end{subequations}
where now the single-particle supergraviton states are given by
\begin{subequations}
\label{1p_R_states}
 \begin{align}
 \ket{\psi;k,m,n}   ^{\rm R}&=\tfrac{1}{m!\,n!}(J_{-1}^+)^m (L_{-1}-J_{-1}^3)^n \ket{\psi}_k^{\rm R},\label{1p_R_states1}\\[1ex]
 \ket{\psi;k,m,n,A} ^{\rm R}&=\tfrac{1}{(m-1/2)!\,(n-1/2)!}(J_{-1}^+)^{m-1/2} (L_{-1}-J_{-1}^3)^{n-1/2} G_{-1}^{+,A} \ket{\psi}_k^{\rm R},\label{1p_R_states2}\\[1ex]
 \ket{\psi;k,m,n,12}^{\rm R}&=\tfrac{1}{(m-1)!\,(n-1)!}(J_{-1}^+)^{m-1} (L_{-1}-J_{-1}^3)^{n-1} 
  \left(G_{-1}^{+,1}G_{-1}^{+,2}+\tfrac{1}{2h^{\rm NS}}(L_{-1}-J_{-1}^3)J_{-1}^+\right) \ket{\psi}_k^{\rm R}.\label{1p_R_states3}
 \end{align}
\end{subequations}
The range of $m$ and $n$ is the same as for \eqref{1p_1/8_sugrtn}.
The operators acting on the R ground states~$\ket{\psi}_k^{\rm R}$ have charges that have been shifted and
sign-flipped due to spectral flow.  Coherent superpositions of the
states \eqref{1/8_sgrvtn_R} are realized as superstrata in the bulk, as
we will expand below.

\subsection{General 1/8-BPS states}
\label{ss:general_1/8-BPS_states}

The states we discussed above represent a large class of 1/8-BPS states
that are nicely in correspondence with the multi-particle supergraviton
states around the bulk ${\rm AdS}_3\times S^3$ background. However, they
are \emph{not} the most general 1/8-BPS states.  This is because we used
only the rigid generators, $L_{-1},G_{-1/2}^{\alpha A},J_0^i$, to excite
the left-moving sector of the theory.  In the Cardy regime, $N_P\gg N$,
this class of 1/8-BPS states has entropy
$S^{\text{1/8-BPS}}_{\text{supergravitons}}\sim N^{1/2}N_P^{1/4}$
\cite{Shigemori:2019orj}.

The most general 1/8-BPS states are obtained by exciting
the left-moving sector by general modes of the fields of the theory.
In the case of $\cM=T^4$, the fundamental fields of the CFT are free
bosonic and fermionic fields, which we collectively denote by
$X,\Psi,\tilde{\Psi}$.\footnote{These fields have symmetry indices as
$X^{A\dot{A}}(z,\zb),\Psi^{\alpha\dot{A}}(z),\tilde{\Psi}^{\dot{\alpha}\dot{A}}(\zb)$
and each has four components on each strand \cite{David:2002wn,
Avery:2010qw}.} On a strand of length $k$, these fields have left-moving
modes $\alpha_{-{n\over k}},\Psi_{-{n+1/2\over k}}$ (in the NS sector),
and the most general states can be obtained by exciting them in an
arbitrary way on all strands, except that the symmetric orbifold
symmetry requires $L_0-\tilde{L}_0$ on each strand to be an integer.
For $\cM={\rm K3}$, we can work at the orbifold point of the K3 moduli
space~\cite{Walton:1987bu}, where K3 is an orbifold of $T^4$, and
project out states that are not invariant under the orbifold action as
well as include the twisted sector.  For both $\cM=T^4$ and ${\rm K3}$,
this will give central charge $c=6N$ worth of 1/8-BPS states.
Alternatively, we can think of the action of the Affine generators
$L_{-{n\over k}}, G^{\alpha A}_{-{n+1/2\over k}}, J^i_{-{n\over k}}$. On
each copy of $\cM$, these generate an $SU(1,1|2)_L$ current algebra of
level~1, leading to central charge $c_{SU(1,1|2)_L}^{}={3\over 2}N$ worth of 1/8-BPS states
\cite{Bena:2014qxa}.
So, In the Cardy regime, the most general 1/8-BPS states have entropy
$S^{\text{1/8-BPS}}_{\text{general}}\sim \sqrt{cN_P}\sim
N^{1/2}N_P^{1/2} \gg S^{\text{1/8-BPS}}_{\text{supergravitons}}$.

The above description of general 1/8-BPS states is valid at the orbifold
point of the D1-D5 CFT\@.  If we perturb the CFT away from the orbifold
point, some of those 1/8-BPS states will lift.  Supergraviton
states (namely, superstrata) are expected to remain supersymmetric on
physical grounds, but more general 1/8-BPS states can lift.  Elliptic genus
and its generalization \cite{Dijkgraaf:1996xw, Maldacena:1999bp,
deBoer:1998us} give partial information about the number of states that
lift, but precisely which states lift is a highly non-trivial problem
and a satisfactory understanding has not emerged yet.\footnote{For
recent progress, see \cite{Hampton:2018ygz, Guo:2019ady}.}

\subsection{Phase diagram}

\begin{figure}[htbp]
 \begin{quote}
  \begin{center}
   \includegraphics[height=6cm]{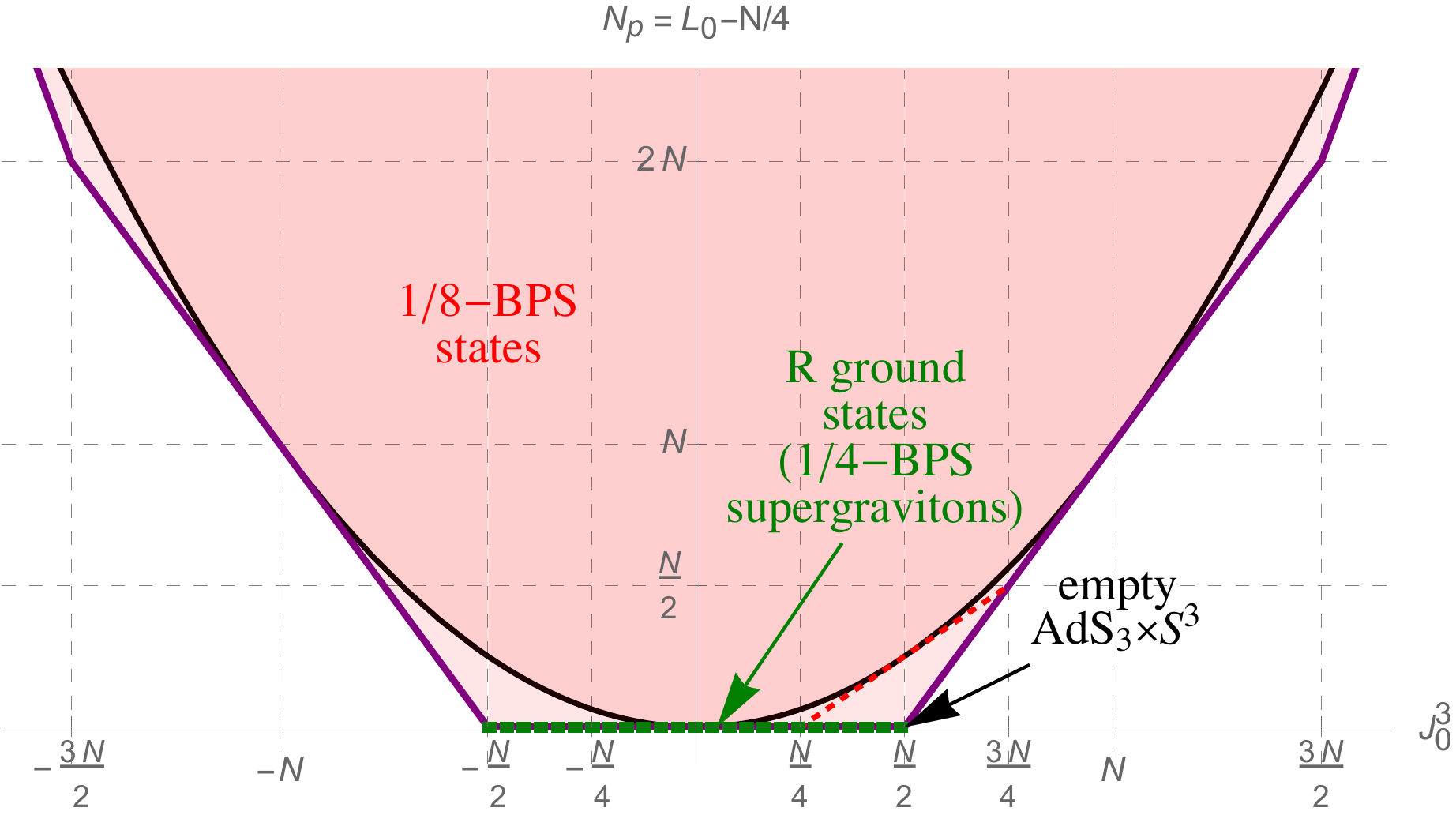} \caption{ The
   $J_0^3$-$N_P$ plane of the D1-D5(-P) system in the R sector.
 \label{fig:phase_diag}}
  \end{center}
 \end{quote}
\end{figure}

The ``phase diagram'' of the states of D1-D5(-P) system on the $J$-$N_P$
plane in the R sector is shown in Figure~\ref{fig:phase_diag}.  Here,
$N_P=L_0-{N/ 4}$.  This is only for the left-moving sector; for
supersymmetry, right-moving sector must be in one of the R ground
states.

States exist only in the region bounded below by the unitarity
bound (the purple polygon in Figure~\ref{fig:phase_diag}).
The empty ${\rm AdS}_3\times S^3$ corresponds to the point
$(J,N_P)=({N/ 2},0)$.  We can think of other states as excitation of
this state.
The 1/4-BPS states are on the interval $J\in [-{N/ 2},{N/ 2}]$, $N_P=0$
(the green dashed line in Figure~\ref{fig:phase_diag}).
The 1/8-BPS states (both supergraviton states and more general states)
have $N_P>0$.  The single-center, 3-charge BMPV black hole exists only
above the parabola $N_P={J^2/ N}$, which is finitely away from the empty
${\rm AdS}_3\times S^3$ point.
%

\section{Supergravity setup}
\label{sec:sugra_setup}

\subsection{The 10-dimensional solution ansatz}

The D1-D5-P black hole is a configuration in type IIB string theory and
is 1/8 BPS, meaning that it preserves 8 supercharges out of the 32
supercharges in 10 dimensions.  Every microstate of the D1-D5-P black
hole must preserve the same supersymmetry.  The most general solutions
of type IIB supergravity that preserve the same 1/8 of supersymmetry and
preserve the symmetry of the internal manifold $\cM$ were studied in
\cite{Giusto:2013rxa, Giusto:2013bda}.
By preserving the symmetry of~$\cM$ we mean that all fields are
independent of the coordinates of $\cM$ and all form fields have legs of
the form $dx^\mu \wedge dx^\nu \wedge \cdots$ or
$\mathrm{vol}(\cM)\wedge dx^\mu \wedge dx^\nu \wedge \cdots$, where
$\mathrm{vol}(\cM)$ is the volume 4-form of $\cM$ and $\mu,\nu,\dots$ are
not along $\cM$.\footnote{For supersymmetric solutions that do not preserve this
symmetry, see \cite{Bakhshaei:2018vux}.}  So, we can forget
about the internal manifold~$\cM$, except for its overall volume, and
consider the remaining six directions.

Preserving the same supersymmetry as the D1-D5-P black hole implies that
the solution must have a null Killing vector,\footnote{There are also
supersymmetric solutions with a timelike Killing vector, but they are
not relevant for the microstates of the D1-D5-P black hole whose
Killing spinor squares to a null Killing vector
\cite{Bossard:2019ajg}.
} which
is chosen to be the direction of a coordinate $u$, and all fields must
be independent of $u$.  The null Killing vector introduce a $2+4$ split
of the six directions and it is natural to introduce a second retarded
time coordinate $v$ and a four-dimensional spatial base $\cB$ with
coordinates $x^m$, $m=1,2,3,4$.  All fields, including the metric of the
base, are independent of $u$ but can depend on $v$ and $x^m$.

The ten-dimensional fields are given by \cite[Appendix E]{Giusto:2013bda}:
\begin{subequations}\label{ansatzSummary}
 \begin{align}
d s^2_{10} & = \sqrt{Z_1 Z_2\over \cP}\,ds_6^2 + \sqrt{\frac{Z_1}{Z_2}}\,ds^2(\cM),\label{10dmetric}\\
ds_6^2  &=-\frac{2}{\sqrt{\cP}}(d v+\beta)\Big[d u+\omega + \frac{\mathcal{F}}{2}(d v+\beta)\Big]+\sqrt{\cP}\,d s^2(\cB),\\
e^{2\Phi}&={Z_1^2\over \cP} ,\qquad
B_2= -\frac{Z_4}{\cP}(d u+\omega) \wedge(d v+\beta)+ a_4 \wedge  (d v+\beta) + \delta_2, \label{Bform}\\ 
C_0&=\frac{Z_4}{Z_1} ,\qquad
C_2 = -{Z_2 \over \cP}(d u+\omega) \wedge(d v+\beta)+ a_1 \wedge  (d v+\beta) + \gamma_2,\\ 
C_4 &= \frac{Z_4}{Z_2} \mathrm{vol}(\cM) - \frac{Z_4}{\cP}\gamma_2\wedge (d u+\omega) \wedge(d v+\beta)+x_3\wedge(d v + \beta) ,\\
C_6 &=\mathrm{vol}(\cM) \wedge \left[ -{Z_1\over \cP}(d u+\omega) \wedge(d v+\beta)+ a_2 \wedge  (d v+\beta) + \gamma_1\right] 
\label{jtta10Apr18} 
\end{align}
\end{subequations}
where
\begin{align}
\cP \equiv Z_1\,Z_2 - Z_4^2.
\end{align}
Here, $ds_{10}^2$ is the string-frame metric of the ten-dimensional
spacetime, $ds^2(\cM)$ is the metric of the internal manifold, and
$ds_6^2$ is the Einstein-frame metric of the six-dimensional spacetime
which involves $u,v$ and the 4-dimensional manifold $\mathcal{B}$ whose
(possibly ambi-polar) metric is
\begin{align}
 ds^2(\cB)=h_{mn}(x,v)dx^m dx^n.
\end{align}
The solution ansatz \eqref{ansatzSummary} contains various quantities:
$Z_{1,2,4}, \cF$ are scalars, $\beta, \omega, a_{1,2,4}$ are 1-forms, $\gamma_{1,2}, \delta_2$ are 2-forms, and
$x_3$ is a 3-form, all on $\mathcal{B}$, and can in general depend on
$v$ but not $u$.  The RR potentials $C_p$ can have extra terms
proportional to a four-form $\cC$ on $\mathcal{B}$, but it has been set
to zero by using an appropriate gauge 
\cite{Giusto:2013rxa}.

The diffeomorphism that preserves the form of the solution ansatz
\eqref{ansatzSummary} is
\begin{align}
 v\to v+V(x),\qquad u\to u+U(x,v),\label{u,v_diffeo}
\end{align}
which induces the following gauge transformation:
\begin{align}
\beta \to \beta - \tilde{d}V,\qquad
 \cF\to \cF-2\dot{U},\qquad
 \omega \to \omega - \tilde{d}U+\dot U \beta.\label{gauge_trans}
\end{align}
Here $\dot{~}\equiv \partial_v$ and we introduced the exterior
derivative restricted to $\cB$,
\begin{align}
  \tilde{d}\equiv dx^m\partial_m.
\end{align}
It will be useful to introduce a differential operator $\cD$  defined
by
\begin{align}
 \cD\equiv \tilde{d}-\beta\wedge \p_v,
\end{align}
which is invariant under the gauge transformation \eqref{u,v_diffeo} and
\eqref{gauge_trans} provided that everything is $u$-independent.  The
full exterior derivative\footnote{The six-dimensional exterior
derivative acting on $u,v,x^m$, although nothing depends on $u$.} can be written as
\begin{align}
 d=\cD+(dv+\beta)\wedge \partial_v.
\end{align}

The $u,v$ coordinates are related to the time coordinate $t$ and the
coordinate $y$ parametrizing the $S^1$ with periodicity $2\pi R_y$, on
which D1- and D5-branes are wrapped.  In view of the gauge symmetry
\eqref{gauge_trans}, the identification is not unique but, in the
current article, we take it to be\footnote{For example, when one relates
6D and 5D solutions, other choices are more convenient; see
\cite{Bena:2017geu}. }
\begin{align}\label{sptvw}
u = \frac{1}{\sqrt{2}}(t-y), \qquad v = \frac{1}{\sqrt{2}}(t+y).
\end{align}
Ignoring the $u$ direction on which nothing depend, we can regard $v$ as
the coordinate of the compact $S^1$ direction.

The quantities $a_{1,2,4},\gamma_{1,2},\delta_2,x_3$ that appear in the
NSNS and RR potentials in \eqref{ansatzSummary} are not invariant under
the gauge symmetry of these potentials.  Gauge-invariant combinations are
\cite{Giusto:2013bda, Bena:2015bea}
\begin{subequations}
\label{def_Theta}
\begin{gather}
 \begin{aligned}
 \Theta_1&\equiv \cD a_1+\dot{\gamma}_2-\dot{\beta}\wedge a_1,~&
 \Theta_2&\equiv \cD a_2+\dot{\gamma}_1-\dot{\beta}\wedge a_2,~&
 \Theta_4&\equiv \cD a_4+\dot{\delta}_2-\dot{\beta}\wedge a_4,
  \\
 \Sigma_1&\equiv \cD \gamma_2 - a_1 \wedge \cD \beta,&
 \Sigma_2&\equiv \cD \gamma_1 - a_2 \wedge \cD \beta,&
 \Sigma_4&\equiv \cD \delta_2 - a_4 \wedge \cD \beta,
 \end{aligned}
 \\
  \Xi_4\equiv \cD x_3-\dot{\beta}\wedge x_3-\Theta_4\wedge \gamma_2
  +a_1\wedge \Sigma_4,
\end{gather}
\end{subequations}%
where $\Theta_I$ are 2-forms and $\Xi_4$ is a 4-form,
and the field strengths can be written in terms of these quantities (see
section~\ref{ss:field_strengths}
 for the explicit expressions).  From this
definition \eqref{def_Theta} and the relations $\cD^2=-(\cD\beta)\wedge
\p_v, \dot{\cD}=-\dot{\beta}\wedge \p_v$, we can show that the following
relations hold between $\Theta_I$ and $\Sigma_I$:
\begin{align}
 \cD\Sigma_I&=- \Theta_I\wedge \cD\beta,\qquad
 \p_v(\Sigma_I+\beta\wedge \Theta_I)=\tilde{d}\Theta_I.
\label{ThetaXiIdentities}
\end{align}
The scalars $Z_1$, $Z_2$, and $Z_4$ can be regarded as the electrostatic
potentials sourced by D1($v$), D5($v\cM$), and F1($v$), respectively,
where D1($v$) means D1-branes extending along $v$.  The 2-forms
$\Theta_2$, $\Theta_1$, and $\Theta_4$ can be regarded as the magnetic
fields sourced by D1($\cC$), D5($\cC\cM$), and F1($\cC$), respectively,
where $\cC$ is a curve in the base $\cB$.

\subsection{The zeroth layer}

The BPS equations satisfied by the ansatz quantities can be organized
in three layers.  The zeroth layer is about the base space $\cB$ and the
1-form $\beta$ on it.  The base $\cB$ must be an almost hyper-K\"ahler
space with three anti-self-dual 2-forms
\begin{align}
 J^{(A)}\equiv {1\over 2}J^{(A)}_{mn}dx^m \wedge dx^n,\qquad
 *_4 J^{(A)} = - J^{(A)}
\end{align}
where $A=1,2,3$ and
$*_4$ is the Hodge star with respect to the metric $ds^2(\cB)$.
The 2-forms satisfy
the quaternionic relation
\begin{align}
 J^{(A)m}{}_p J^{(B)p}{}_n=\epsilon^{ABC}J^{(C)m}{}_n-\delta^{AB}\delta^m_n\label{khrn4Aug19}
\end{align}
where the indices are raised and lowered using $h_{mn}$ and its inverse
$h^{mn}$. Unlike in hyper-K\"ahler spaces, these 2-forms are not closed;
instead, they are required to satisfy
\begin{align}
 \tilde{d}J^{(A)}=\partial_v(\beta\wedge J^{(A)}).\label{gwhj3Aug19}
\end{align}
Furthermore, $\beta$ must satisfy
\begin{align}
 \cD \beta=*_4 \cD\beta.\label{khru4Aug19}
\end{align}
The integrability condition for \eqref{gwhj3Aug19}, obtained by acting
on it with $\tilde{d}$, is $\p_v(\cD\beta\wedge J^{(A)})=0$, which is
guaranteed to hold because $\cD \beta$ is self-dual and $J^{(A)}$ is
anti-self-dual. 

One quantity that is defined by the data of the zeroth
layer is the anti-self-dual 2-form
\begin{align}
 \psi\equiv {1\over 8}\epsilon^{ABC}J^{(A)mn}\dot{J}^{(B)}{}_{mn}J^{(C)},
\label{def_psi}
\end{align}
which will show up in higher layers.

In the zeroth layer, we must find almost complex structures $J^{(A)}$
and a 1-form $\beta$ which in general depend on $v$ and satisfy the
non-linear conditions \eqref{khrn4Aug19}--\eqref{khru4Aug19}. If we solve the zeroth layer, the remaining two
layers can be written as linear differential equations on $\cB$.  In
practice, in most of the explicit superstratum solutions in the
literature, it is assumed that the $\cB$ is flat $\bbR^4$
or a Gibbons-Hawking space
\cite{Bena:2017geu, Tyukov:2018ypq, Walker:2019ntz},
and that $\beta$ is
independent of $v$.

\subsection{The first layer}

The first-layer equations determine the scalars $Z_I$ and the flux
forms $\Theta_I, \Sigma_I$.  They must satisfy \cite{Giusto:2013bda} the
following linear differential equations
\begin{align}
  *_4(\cD Z_1+\dot{\beta} Z_1)&=\Sigma_2,&\quad
 *_4(\cD Z_2+\dot{\beta} Z_2)&=\Sigma_1,&\quad
 *_4(\cD Z_4+\dot{\beta} Z_4)&=\Sigma_4,\label{iql6Aug19}
\end{align}
and duality relations
\begin{align}
 (1-*_4)\Theta_2=2Z_1\psi, \qquad
 (1-*_4)\Theta_1=2Z_2\psi, \qquad
 (1-*_4)\Theta_4=2Z_4\psi.\label{Theta_psi_duality}
\end{align}
Another condition is
\begin{align}
\Xi_4=Z_2^2\,\partial_v\Bigl({Z_4\over Z_2}\Bigr)\,{*_4 1}.
\end{align}

By acting with $\p_v$ and $\cD$ on \eqref{iql6Aug19} and using the
identities \eqref{ThetaXiIdentities}, we can derive equations involving
only $Z_I,\Theta_I$:
\begin{subequations}\label{sfd6Aug19}
 \begin{align}
  \p_v [*_4 (\cD Z_1+\dot{\beta}Z_1)+\beta\wedge \Theta_2]&=\tilde{d}\Theta_2,\\
  \p_v [*_4 (\cD Z_2+\dot{\beta}Z_2)+\beta\wedge \Theta_1]&=\tilde{d}\Theta_1,\\
  \p_v [*_4 (\cD Z_4+\dot{\beta}Z_4)+\beta\wedge \Theta_4]&=\tilde{d}\Theta_4
 \end{align}
\end{subequations}
and
\begin{subequations} 
\label{L1-integrability}
 \begin{align}
 \cD *_4 (\cD Z_1+\dot{\beta}Z_1)&=-\Theta_2 \wedge \cD\beta,\\
 \cD *_4 (\cD Z_2+\dot{\beta}Z_2)&=-\Theta_1 \wedge \cD\beta,\\
 \cD *_4 (\cD Z_4+\dot{\beta}Z_4)&=-\Theta_4 \wedge \cD\beta.
 \end{align}
\end{subequations}
Eqs.~\eqref{L1-integrability} can be regarded as the integrability condition for
\eqref{iql6Aug19} or \eqref{sfd6Aug19}.

Given the solution of the zeroth layer, Eqs.~\eqref{Theta_psi_duality} and
\eqref{sfd6Aug19}  give a system of linear equations defined on~$\cB$
which can be solved to determine $Z_I,\Theta_I$.  Each line of
\eqref{sfd6Aug19} contains four equations, which can be used to find
four independent components of $Z_I,\Theta_I$ in view of the
duality relation~\eqref{Theta_psi_duality}.

\subsection{The second layer}

Given the solution to the first-layer equations, the second-layer
equations give a system of linear equations for $\omega,\cF$ with 
sources quadratic in the  first-layer fields
\cite{Giusto:2013bda}:
\begin{subequations} 
  \label{L2eqs}
 \begin{align}
  \label{L2-1}
 (1+*_4)\cD\omega+\cF\, \cD\beta  
 =Z_1 \Theta_1+Z_2 \Theta_2-2Z_4\Theta_4-2(Z_1 Z_2-Z_4^2)\psi,
 \end{align}
 \begin{align}
  \label{L2-2}
 *_4 \cD *_4 L &+ 2\dot{\beta}_m L^m -*_4(\psi\wedge \cD\omega)
\notag\\
 & =
  -{1\over 4}(Z_1 Z_2-Z_4^2)\dot{h}^{mn} \dot{h}_{mn}
 +{1\over 2}\p_v[(Z_1 Z_2-Z_4^2)h^{mn} \dot{h}_{mn}]\notag\\
 &\quad\,
 +(\dot{Z}_1\dot{Z}_2-\dot{Z}_4^2)
 +(Z_1\ddot{Z}_2+Z_2\ddot{Z}_1-2Z_4\ddot{Z}_4)\notag\\
 &\quad\,
 -{1\over 2}*_4\Bigl[
 (\Theta_1-Z_2\psi)\wedge (\Theta_2-Z_1\psi)
 - (\Theta_4-Z_4\psi)\wedge (\Theta_4-Z_4\psi)\notag\\
 &\qquad\qquad\qquad
 +(Z_1 Z_2-Z_4^2)\psi\wedge \psi
 \Bigr],
 \end{align}
\end{subequations}
where
\begin{align}
 L\equiv \dot{\omega}+{\cF\over 2}\dot{\beta}-{1\over 2}\cD\cF.
  \label{def_L}
\end{align}

\subsection{Field strengths}
\label{ss:field_strengths}

Using the relations above, the NSNS and RR field strengths can be
written solely in terms of $\beta,Z_I,\Theta_I,\omega$.  The explicit
expression for the NSNS field strength is
\cite{Giusto:2013rxa}
\begin{align}
 H_3=dB_2
 &=
 -d\left[ {Z_4\over \cP}(du+\omega)\wedge (dv+\beta)\right]
 +(dv+\beta)\wedge \Theta_4 + *_4(\cD Z_4+\dot{\beta}Z_4).
\end{align}
The RR field strengths are defined by $G_{p+1}=dC_p-H_3\wedge C_{p-2}$.
In the present case, their explicit from can be conveniently written in
terms of $F_p,\tilde{F}_p$ defined by
\begin{align}
 G=\sum_{p=1,3,4,5,7,9}G_p
 =: F_1+F_3+F_5+(\tilde{F}_1+\tilde{F}_3+\tilde{F}_5)\wedge \mathrm{vol}(\cM).
\end{align}
The explicit expressions for $F_p$ are
\begin{align}
 F_1&=\cD\left({Z_4\over Z_1}\right)
 +(dv+\beta)\,\partial_v\left({Z_4\over Z_1}\right),\\
 F_3&=-(du+\omega)\wedge (dv+\beta) \wedge 
 \left[
 \cD\left({1\over Z_1}\right)-{1\over Z_1}\dot{\beta}
 +{Z_4\over Z_1}\cD\left({Z_4\over \cP}\right)
 \right]\notag\\
 &\qquad
 +(dv+\beta)\wedge \left(\Theta_1-{Z_4\over Z_1}\Theta_4-{1\over Z_1}\cD\omega\right)
 +{1\over Z_1}(du+\omega)\wedge \cD\beta
 \notag\\
 &\qquad
 +*_4\left[
 (\cD Z_2+\dot{\beta}Z_2)
 -{Z_4\over Z_1}
 (\cD Z_4+\dot{\beta}Z_4)
 \right],\\
 F_5&=
 {1\over \cP}(du+\omega)\wedge (dv+\beta) \wedge *_4\left[
 Z_2(\cD Z_4+\dot{\beta}Z_4)
 -Z_4(\cD Z_2+\dot{\beta}Z_2)
 \right]\notag\\
 &\qquad
 +(dv+\beta)\wedge Z_2^2\partial_v\left({Z_4\over Z_2}\right)\,{*_4 1}.
\end{align}
$\tilde{F}_p$ can be obtained from $F_p$ by setting $Z_1\leftrightarrow
Z_2$, $\Theta_1\leftrightarrow \Theta_2$.

\subsection{A covariant form of BPS equations}

It is possible to write the above BPS equations in a more concise form \cite[Appendix A]{Bena:2017geu}.
Define the matrix
\begin{align}
 C^{IJ}=
 \begin{pmatrix}
  0 & 1 & 0  \\
  1 & 0 & 0  \\
  0 & 0 & -2 \\
 \end{pmatrix}, \qquad
 C_{IJ}=
 \begin{pmatrix}
  0 & 1 & 0  \\
  1 & 0 & 0  \\
  0 & 0 & -\thalf \\
 \end{pmatrix}, \qquad
 I,J=1,2,4,
\end{align}
which has origin in an M-theory frame as intersection numbers of
2-cycles \cite{Giusto:2012gt, Bena:2015bea}. The matrices $C_{IJ}$ and $C^{IJ}$ are
inverse matrices of each other.  Furthermore, define
$\sfa^I,\sfgamma^I,\sfZ_I,\sfTheta^I,\sfSigma^I$ by
\begin{align}
\begin{gathered}
  \sfa^1\equiv a_1,\quad \sfa^2\equiv a_2,\quad \sfa^4\equiv 2a_4,\qquad
 \sfgamma^1\equiv \gamma_2,\quad \sfgamma^2\equiv \gamma_1,\quad \sfgamma^4\equiv 2 \delta_2,\\
 \sfZ_1\equiv Z_1,\quad \sfZ_2\equiv Z_2,\quad \sfZ_4\equiv -Z_4,\qquad
 \sfTheta^1\equiv \Theta_1,\quad \sfTheta^2\equiv \Theta_2,\quad \sfTheta^4\equiv 2\Theta_4,\\
 \sfSigma^1\equiv \Sigma_1,\quad \sfSigma^2\equiv \Sigma_2,\quad \sfSigma^4\equiv 2\Sigma_4.
\end{gathered}
\end{align}
We raise
and lower indices using $C^{IJ}$ and $C_{IJ}$; for example,
$\sfZ^I=C^{IJ}\sfZ_J$ and $\sfTheta_I=C_{IJ}\sfTheta^I$.  We define
the inner product by
\begin{align}
 (\sfA,\sfB)\equiv A_I B^I.
\end{align}
Then we can rewrite the BPS equations as follows.

\noindent
Definitions of $\Theta_I,\Sigma_I$, \eqref{def_Theta}:
\begin{align}
 \sfTheta&=\cD \sfa + \dot\sfgamma-\dot{\beta}\wedge \sfa ,\qquad
 \sfSigma=\cD \sfgamma-\sfa\wedge \cD \beta,
\end{align}
The first-layer equations \eqref{iql6Aug19}--\eqref{L1-integrability}:
\begin{gather}
  *_4 (\cD \sfZ+\dot{\beta}\sfZ)= \sfSigma,\qquad
  (1-*_4)\sfTheta=2 \sfZ \psi\\
 \p_v [*_4 (\cD \sfZ+\dot{\beta}\sfZ)+\beta\wedge \sfTheta]=\tilde{d}\sfTheta,\qquad
 \cD *_4 (\cD \sfZ+\dot{\beta}\sfZ)=-\sfTheta \wedge \cD\beta
\end{gather}
The second-layer equations \eqref{L2eqs}:
\begin{subequations} 
 \begin{align}
 (1+*_4)\cD\omega+\cF\, \cD\beta  
 &=(\sfZ,\sfTheta)-(\sfZ,\sfZ)\psi\\
 *_4 \cD *_4 L + 2\dot{\beta}_m L^m -*_4(\psi\wedge \cD\omega)
 & =
 -{1\over 8}(\sfZ,\sfZ)\dot{h}^{mn} \dot{h}_{mn}
 +{1\over 4}\p_v[(\sfZ,\sfZ)h^{mn} \dot{h}_{mn}]\notag\\
 &\quad +{1\over 2}(\dot\sfZ,\dot\sfZ) +(\sfZ,\ddot{\sfZ})\notag\\
 &\quad 
 -{1\over 4}*_4\Bigl[
 (\sfTheta-\sfZ \psi   \ \overset{\wedge }{,}\ \sfTheta-\sfZ \psi)
 +(\sfZ,\sfZ)\psi\wedge \psi \Bigr]
 \end{align}
\end{subequations}

The 10-dimensional solution \eqref{ansatzSummary}, having no dependence
in the internal manifold $\cM$, can also be studied within 6-dimensional
supergravity.  In $d=6$, $\cN=(1,0)$ supergravity~\cite{Nishino:1984gk,
Nishino:1986dc}, a graviton multiplet consists of a graviton
$g_{\mu\nu}$, a left-handed symplectic Majorana-Weyl gravitino
$\psi_\mu$, and a tensor gauge field $B^+_{\mu \nu}$ with self-dual
dressed field-strength.  A tensor multiplet consists of a two-form
$B^-_{\mu\nu}$ with anti-self-dual dressed field-strength, a
right-handed symplectic Majorana-Weyl fermion $\chi$, and a scalar field
$\varphi$.  Classification of supersymmetric solutions in minimal
$d=6,\cN=(1,0)$ supergravity was done in \cite{Gutowski:2003rg} and
later extended to include other multiplets in \cite{Cariglia:2004kk,
Lam:2018jln, Cano:2019gqm}.  Classification of supersymmetric solutions
in $d=6,\cN=(2,0)$ supergravity was carried out in
\cite{Bossard:2019ajg}.  The theory without $(Z_4,\Theta_4)$ corresponds
to minimal $\cN=(1,0)$ supergravity plus a tensor multiplet, and
including $(Z_4,\Theta_4)$ means to add another tensor multiplet
\cite{Kanitscheider:2007wq}.  The index $I$ above corresponds to the
label for the (self-dual and anti-self-dual) tensor gauge fields
$B^{\pm}_{\mu\nu}$.

\subsection{$v$-independent case}

In the above, we wrote down the BPS equations in the general case where
the base space metric $ds^2(\cB)$ and the 1-form $\beta$ are
$v$-dependent.  To consider general microstate geometries, such general
base space is unavoidable.  However, because of technical limitation, it
is normally assumed that $ds^2(\cB)$ and $\beta$ are independent of $v$.
This certainly restricts the class of microstate geometries, but there
are superstrata with such a base whose entropy scales the same as the
general superstrata ($S\sim N^{1/2}N_P^{1/4}$); see section~\ref{ss:00_superstrata}.

Here, we assume that the zeroth-layer ansatz quantities, namely the base
space metric $ds^2(\cB)$, the 1-form $\beta$, and 2-forms $J^{(A)}$, do
not depend on~$v$ and write down the form of the BPS equations we
introduced above.
In the zeroth layer, the complex structures $J^{(A)}$ are closed and
therefore the base space $\cB$ becomes hyper-K\"ahler.  The condition on
$\beta$ is that it is self-dual,
\begin{align}
 d\beta=*_4 d\beta.\label{db_SD_v-indep}
\end{align}
Also, the anti-self-dual 2-form $\psi$ defined in \eqref{def_psi} vanishes.

Under the above assumptions, the first-layer equations
\eqref{Theta_psi_duality}--\eqref{L1-integrability} become
\begin{subequations}\label{L1eqs'}
\begin{align}
*_4\cD\dot{Z}_1 = \cD \Theta_2, \qquad \cD *_4 \cD Z_1 = - \Theta_2 \wedge d\beta,\qquad \Theta_2 = *_4 \Theta_2,\label{L1eqs-1'}\\
*_4\cD\dot{Z}_2 = \cD \Theta_1, \qquad \cD *_4 \cD Z_2 = - \Theta_1 \wedge d\beta,\qquad \Theta_1 = *_4 \Theta_1,\label{L1eqs-2'}\\
*_4\cD\dot{Z}_4 = \cD \Theta_4, \qquad \cD *_4 \cD Z_4 = - \Theta_4 \wedge d\beta,\qquad \Theta_4 = *_4 \Theta_4.\label{L1eqs-4'}
\end{align}
\end{subequations}

The second-layer equations \eqref{L2eqs} simplify to
\begin{subequations}\label{L2eqs'}
\begin{align}
(1+*_4)\cD \omega +\cF d\beta &= Z_1 \Theta_1 + Z_2 \Theta_2 - 2 Z_4 \Theta_4,\\
*_4 \cD *_4\left(\dot{\omega}- \frac12 \cD \cF\right) &=
 (\dot{Z}_1\dot{Z}_2-\dot{Z}_4^2)
 +(Z_1\ddot{Z}_2+Z_2\ddot{Z}_1-2Z_4\ddot{Z}_4)
 \notag\\[-1ex]
& \hspace{15ex}
 - \frac12 *_4(\Theta_1\wedge \Theta_2 - \Theta_4 \wedge \Theta_4),
\end{align}
\end{subequations}
where
\begin{align}
 L= \dot{\omega}-{1\over 2}\cD\cF.
\end{align}

\subsection{2-charge microstate geometries}

Let us see how the 2-charge microstate geometries (Lunin-Mathur geometries),
which are dual to the 1/4-BPS supergraviton states
\eqref{1/4_sgrvtn_R}, are described in the supergravity setup above.
The Lunin-Mathur geometries that respect the symmetry of $\cM$ are
parametrized by profile functions $g_A(\lambda )$ with $A=1,2,3,4=:i$ and
$A=5$.\footnote{The $A\ge 6$ components \cite{Lunin:2002iz,
Kanitscheider:2007wq} break the symmetry of $\cM$.}  Given such a
profile, the ansatz data are given by \cite{Lunin:2001jy, Lunin:2002iz,
Kanitscheider:2007wq, Bena:2015bea}
\begin{subequations}\label{LM_geom_general}
\begin{align}
 Z_1 &= 1 + \frac{Q_5}{L} \int_0^{L} d\lambda \frac{|\partial_\lambda g_i(\lambda )|^2+|\partial_\lambda g_5(\lambda )|^2}{|x_i -g_i(\lambda )|^2} ,\qquad
  Z_4 = - \frac{Q_5}{L} \int_0^{L} d\lambda \frac{\partial_\lambda g_5(\lambda )}{|x_i -g_i(\lambda )|^2} ,\\\label{Z1profile}
 Z_2 &= 1 + \frac{Q_5}{L} \int_0^{L} \frac{d\lambda }{|x_i -g_i(\lambda )|^2}, \qquad d\gamma_2 = *_4 d Z_2,\qquad d\delta_2 = *_4 d Z_4,\\
 A &= - \frac{Q_5}{L} dx^j \int_0^{L} d\lambda  \frac{\partial_\lambda g_j(\lambda )}{|x_i -g_i(\lambda )|^2} , \qquad dB = - {*_4 dA},\qquad ds^2(\cB) = dx^i dx^i, \label{general2chg-A,B}\\
 \beta &= \frac{-A+B}{\sqrt{2}},\qquad\omega = \frac{-A-B}{\sqrt{2}},\qquad
 \Theta_I=\cF=a_{1,4}=x_3=0.
\end{align}
\end{subequations}
The base $\cB$ is always flat $\bbR^4$ with coordinates $x^i$, and 
$*_4$ is the Hodge dual with respect to its flat metric $ds^2_4=dx^i dx^i$. 
The functions
$g_A(\lambda )$ are periodic with period $L=2\pi  Q_5/ R_y$.  
The D1 charge is given by
\begin{align}
 Q_1={Q_5\over L}\int_0^L d\lambda \bigl(|\partial_\lambda g_i(\lambda )|^2+|\partial_\lambda g_5(\lambda )|^2\bigr).\label{Q1_and_Q5}
\end{align}
The quantities $Q_1$, $Q_5$ are related to the quantized D1 and D5
numbers $N_1$, $N_5$ by
\begin{align}
Q_1 = \frac{N_1 g_s \alpha'^3}{v_4}\,,\qquad Q_5 = N_5 g_s \alpha',
\label{Q1Q5_n1n5}
\end{align}
 where $(2\pi)^4 v_4$ is the coordinate volume of $\cM$.

If we drop ``$1$'' from $Z_{1,2}$ in \eqref{LM_geom_general}, the
geometry becomes asymptotically ${\rm AdS}_3$. The AdS/CFT dictionary
between the profile~$g_A(\lambda )$ and the 1/4-BPS supergraviton states
\eqref{1/4_sgrvtn_R} is that the Fourier coefficient with
mode~$k$ in $g_A$ is related to the excitation number $N^\psi_k$ as
follows:\footnote{The absolute value square of the Fourier coefficient
is proportional to $N^\psi_k$ with a non-trivial coefficient. For the
precise map, see, {\it e.g.}, \cite{Giusto:2019qig}.}
\begin{align}
g_1\pm ig_2 ~\leftrightarrow~ N^{\pm\pm}_{k},\qquad\quad
g_3\pm ig_4 ~\leftrightarrow~ N^{\pm\mp}_{k},\qquad\quad
g_5         ~\leftrightarrow~ N^{00}_{k}.
\label{holog_dict_2-chg}
\end{align}
$N^\psi_k$ for other species $\ket{\psi}$ are not turned on, because they would break
the symmetry of $\cM$.

\subsubsection{Empty ${\rm AdS}_3\times S^3$}
\label{sss:empty_AdS3xS3}

The simplest example is the circular profile in the 1-2 plane:
\begin{align}
 g_1+ig_2=a e^{2\pi i \lambda/L},\qquad g_3+ig_4=g_5=0,\label{profile_emptyAdS3xS3}
\end{align}
where $a>0$ is a constant.
According to the dictionary \eqref{holog_dict_2-chg}, this case is dual to
the following RR ground state:
\begin{align}
 [\ket{++}_1]^N,\qquad
 J_L=J_R={N\over 2}, \qquad N_P=0.
\label{emptyAdS3_cft_state}
\end{align}

To write down the ansatz data, it is convenient to write the flat metric
for the base $\bbR^4$ in the following form:
\begin{align}
\label{base_metric_flat}
 d s^2(\cB) &= \Sigma\, \Bigl(\frac{d r^2}{r^2+a^2}+ d\theta^2\Bigr)+(r^2+a^2)\sin^2\theta\,d\phi^2+r^2 \cos^2\theta\,d\psi^2,\\
 \Sigma&\equiv r^2+a^2 \cos^2\theta
.
\end{align}
The relation to the Cartesian coordinates $x^i$  is
\begin{align}
 x^1+ix^2=\sqrt{r^2+a^2}\sin\theta\,e^{i\phi},\qquad
 x^3+ix^4=r\cos\theta\,e^{i\psi}.
\end{align}
In this coordinate system, some of the ansatz data are
\begin{subequations} 
 \label{ansatz_data_empty_AdS3xS3}
 \begin{align}
 Z_1&={R_y^2a^2\over Q_5\Sigma},\qquad
 Z_2={Q_5\over \Sigma},\qquad
 Z_4=0,\qquad \Theta_I=0,
 \label{emptyAdS3_Z_I}
 \\
 \beta&=  \frac{R_y a^2}{\sqrt{2}\,\Sigma}(\sin^2\theta\, d\phi - \cos^2\theta\,d\psi)\equiv \beta_0,
 \label{beta0}
 \\
 \omega&= \frac{R_y a^2}{\sqrt{2}\,\Sigma}(\sin^2\theta\, d\phi + \cos^2\theta\,d\psi)\equiv \omega_0,\qquad \cF=0.
 \label{omega0}
 \end{align}
\end{subequations}
In \eqref{emptyAdS3_Z_I}, we have dropped ``1'' in
\eqref{LM_geom_general}, so the geometry is asymptotically ${\rm
AdS}_3$.  Eq.~\eqref{Q1_and_Q5} relates the parameters $a,R_y,Q_1,Q_5$
as
\begin{align}
 a^2={Q_1 Q_5\over R_y^2}.\label{aQR_y}
\end{align}


This geometry is empty global ${\rm AdS}_3\times S^3$, as we can see by
doing the coordinate transformation
\begin{align}
 \phit
 =\phi-{t\over R_y}
 ,\qquad
 \psit
 &=\psi-{y\over R_y}
 \label{bulk_spectr_flow}
\end{align}
after which the 6D metric becomes
\begin{align}
 ds^2_{6}
 &=\sqrt{Q_1Q_5}
 \left(-{r^2+a^2\over a^2 R_y^2}dt^2+{r^2\over a^2 R_y^2}dy^2
 +{dr^2\over r^2+a^2}
 +d\theta^2+\sin^2\theta\, d\phit^2+\cos^2\theta\, d\psit^2\right).
\end{align}
This is indeed global AdS$_3\times S^3$ with radius
$\cR=(Q_1Q_5)^{1/4}$.  This is consistent with the fact that the R ground state
\eqref{emptyAdS3_cft_state} is mapped via \eqref{NS-R_map_example} into the
unique NS vacuum $[\ket{--}_1^{\rm NS}]^N$.  In the dual CFT, the
coordinate transformation \eqref{bulk_spectr_flow} corresponds to the
spectral flow transformation between the R and NS sectors.

\subsubsection{$Z_4$ excitation}
\label{ss:LM_Z4}

A slightly more non-trivial example is given by the profile
\begin{align}
 g_1+ig_2=ae^{2\pi i \lambda /L},\qquad g_3+ig_4=0,\qquad
 g_5=-{b\over k}\sin{2\pi k \lambda \over L}
\label{circularLM+g5}
\end{align}
which is dual to the state
\begin{align}
 [\ket{++}_1]^{N_0} 
 [\ket{00}_k]^{N^{00}_k},\qquad
 N_0+k N^{00}_k=N.\label{Z4exc_state}
\end{align}
The ansatz data are,  in the 
coordinate system \eqref{base_metric_flat},
\begin{align}
 \label{Z4_excitaion_ansatz_data}
\begin{split}
 Z_1 &= \frac{R^2}{Q_5}\Bigl[\frac{a^2+{b^2/ 2}}{\Sigma} 
 + b^2 a^{2k} \frac{\sin^{2k}\theta\,\cos(2k\phi)}{2(r^2+a^2)^k\,\Sigma}\Bigr],
 \qquad Z_2 = \frac{Q_5}{\Sigma}\,, \\
 Z_4&=R b a^k \frac{\sin^k\theta\,\cos(k\phi)}{(r^2+a^2)^{k/2}\,\Sigma},\qquad
 \beta = \beta_0,\qquad \omega= \omega_0 ,\qquad
 \cF=\Theta_I=0.
\end{split}
\end{align}
The relation \eqref{aQR_y} is modified to
\begin{align}
 a^2+{b^2\over 2}={Q_1 Q_5\over R_y^2}.
\end{align}
Because $a^2\propto N_0,b^2\propto N^{00}_k$, this relation is the bulk
dual of the strand budget constraint given by the second equation of
\eqref{Z4exc_state}.

\section{Superstrata}
\label{sec:superstrata}

In section \ref{sec:CFT}, we discussed states that correspond to
populating ${\rm AdS}_3\times S^3$ with ``supergravitons''.  For the
1/4-BPS supergraviton gas state \eqref{1/4_sgrvtn_R}, when the number of
excited supergravitons is macroscopic, namely, when $N^\psi_k=\cO(N)$
(excluding $N^{++}_1$, which corresponds to the ${\rm AdS}_3\times S^3$
background), the bulk spacetime gets backreacted and is given by the
Lunin-Mathur geometry \eqref{LM_geom_general}.  Similarly, for 
the 1/8-BPS supergraviton gas state \eqref{1/8_sgrvtn_R}, when the
excitation number $N^\psi_{k,m,n,f}$ is $\cO(N)$, the bulk spacetime
must become a backreacted geometry, which can be described within the
supergravity setup of section \ref{sec:sugra_setup}.  In the original
incarnation, this is nothing but the superstratum.
Here let us review the construction of the superstratum.

\subsection{General remarks}

\medskip

\subsubsection{Coherent states}

The states of the 1/4-BPS supergraviton gas, \eqref{1/4_sgrvtn_R}, is
holographically described by the Lunin-Mathur microstate geometries
\eqref{LM_geom_general}.  However, this is not a precise statement.  The
state~\eqref{1/4_sgrvtn_R} with a fixed distribution $\{N^\psi_k\}$
generally corresponds to a highly quantum situation in the bulk and
cannot be described by a well-defined classical geometry.  Instead, the
state that corresponds to a smooth microstate geometry is a coherent
superposition of the state~\eqref{1/4_sgrvtn_R} with different
distributions $\{N^\psi_k\}$ peaked around an average distribution.
This has been well established for 1/4-BPS supergraviton states and
Lunin-Mathur geometries \cite{Skenderis:2006ah, Kanitscheider:2006zf,
Kanitscheider:2007wq}.  Similarly, for the 1/8-BPS case, the state that
corresponds to a well-defined classical geometry is not
\eqref{1/8_sgrvtn_R} but a coherent state obtained by taking a
superposition of~\eqref{1/8_sgrvtn_R} with different distributions
$\{N^\psi_{k,m,n,f}\}$ \cite{Giusto:2015dfa, Bena:2017xbt}.

That we must use a coherent superposition for a classical configuration
is clear from the coherent state for the harmonic oscillator of quantum
mechanics, but an argument in our context is as follows.  A state like
\eqref{1/8_sgrvtn_R} is an eigenstate of charges such as $L_0,J_L$ and
therefore the vev of operators that have non-vanishing charges will
vanish for the state.  However, the vev is holographically related to
the falloff of bulk fields near the boundary, which generally does not
vanish in microstate geometries.  To resolve this contradiction, one
should take a coherent superposition of states with different values of
$L_0,J_L$.  Another, related way to argue that a superposition is
necessary is as follows \cite{Giusto:2013bda}.  The CFT state
\eqref{1/8_sgrvtn_R} is an eigenstate of the momentum operator
$N_P=L_0-\tilde{L}_0$ and is therefore invariant under translation up to
phase.  On the other hand, in the bulk, the geometry carrying $N_P>0$
involves traveling waves along $v$ and thus is $v$-dependent; namely, it
is not invariant under translation.  This problem is resolved if the
precise CFT dual of a bulk geometry is not~\eqref{1/8_sgrvtn_R} but a
coherent superposition of states with different values of $N_P$.

Specifically, the supergraviton state \eqref{1/8_sgrvtn_R} must more
properly be replaced by the following coherent superposition state
characterized by complex parameters $\{A^\psi_{k,m,n,f}\}$:
\begin{align}
 \Psi(\{A^{\psi}_{k,m,n,f}\})
 &=
 \sideset{}{'}\sum_{\{N^\psi_{k,m,n,f}\}}
 \prod_{\psi,k,m,n,f} (A^\psi_{k,m,n,f})^{N^\psi_{k,m,n,f}}\,
 \Psi(\{N^{\psi}_{k,m,n,f}\})\notag\\
 &=
 \sideset{}{'}\sum_{\{N^\psi_{k,m,n,f}\}}
 \prod_{\psi,k,m,n,f} \Bigl[A^\psi_{k,m,n,f}\ket{\psi;k,m,n,f}\Bigr]^{N^\psi_{k,m,n,f}},
 \label{coh_superposition}
\end{align}
where the sum is restricted to distributions $\{N^\psi_{k,m,n,f}\}$ that
satisfy \eqref{1/8_sgrvtn_R_constraint}.  This state is not normalized.
We restrict to the case where $N^\psi_{k,m,n,f}\neq 0$ only for bosonic
$\ket{\psi;k,m,n,f}$.  In the large $N$ limit, the sum is dominated by a
particular distribution $\{\Nb^\psi_{k,m,n,f}\}$, which can be obtained
by computing the norm $|\Psi(\{A^{\psi}_{k,m,n,f}\})|^2$ and taking its
variation with respect to $\{N^{\psi}_{k,m,n,f}\}$.  Generally, we have
$\Nb^\psi_{k,m,n,f}\propto |A^{\psi}_{k,m,n,f}|^2$, although the detail
depends on the normalization of the state
$\Psi(\{N^{\psi}_{k,m,n,f}\})$.  In the current article, we will not
explicitly use coherent states and loosely talk about the geometry
corresponding to the supergraviton state~\eqref{1/8_sgrvtn_R} specified
by the distribution $\{N^\psi_{k,m,n,f}\}$.  However, strictly speaking,
we should instead use the coherent state \eqref{coh_superposition} whose
average distribution is equal to $\{N^\psi_{k,m,n,f}\}$.  For more
detail of the coherent superposition for 1/8-BPS states, see
\cite{Giusto:2015dfa, Bena:2017xbt}.

\subsubsection{Fixing the base}

On physical grounds, the bulk geometry dual to \eqref{1/8_sgrvtn_R} for
any $\{N^\psi_{k,m,n,f}\}$, or more precisely its coherent state
version, is expected to exist and, being BPS, must be obtainable by
solving the BPS equations of section \ref{sec:sugra_setup}.  They are
superstrata.  However, for constructing such general solutions, we must
confront the non-linear problem in the zeroth layer of finding the
almost hyper-K\"ahler base and the associated 1-form,
$(ds^2(\cB),\beta)$, both $v$-dependent, appropriate for the state.  At
the time of writing, this is an unsolved technical problem, because
little is known about the relevant almost hyper-K\"ahler space, and
because we do not know in general what almost hyper-K\"ahler base to
take for a given CFT state.
Instead, let us assume that there is a set of modes $(\psi,k,m,n,f)\in
\cK$ that correspond to excitations with the \emph{same} particular base.
Namely, no matter what $\{N^\psi_{k,m,n,f}\}$ is for $(\psi,k,m,n,f)\in
\cK$, the bulk geometries have the same zeroth-layer data
$(ds^2(\cB),\beta)$. The relevant CFT state is assumed to take the form
\begin{align}
 \Psi_{\rm bg}\times
 \!\!\!\prod_{(\psi,k,m,n,f)\in \cK}\!\!\!
 \bigl[\,\ket{\psi;k,m,n,f}\,\bigr]^{N^{\psi}_{k,m,n,f}},\qquad
 N_{\rm bg}+
 \!\!\!\sum_{(\psi,k,m,n,f)\in \cK}\!\!\!
 k N^{\psi}_{k,m,n,f}=N.
 \label{gen_state_R_same_base}
\end{align}
Here, $\Psi_{\rm bg}$ is a ``background'' part that is made of strands
of total length $N_{\rm bg}$ and corresponds to the fixed base
$(\cB,\beta)$.  Depending on the number $N^{\psi}_{k,m,n,f}$ of
supergravitons added to the system, the total strand length $N_{\rm bg}$
must change to accommodate them, if $N$ is to be fixed.  Alternatively,
we can fix $\Psi_{\rm bg}$ and change the system size $N$ to accommodate the change in~$N^{\psi}_{k,m,n,f}$.

If this assumption holds, the problem reduces to that of solving the BPS
equations in the first and second layers, defined on a \emph{fixed base
$\cB$}.  In the first layer, because the equations are linear and
homogeneous (sourceless), it must be possible to write the solution as a
sum of modes with arbitrary coefficients. For example, assume that we
have solved the equations for the pair $(Z_4,\Theta_4)$.  Then they can
be written as
\begin{subequations} 
\label{L1_expn_general}
\begin{align}
 Z_4(x,v)     =\sum_\sfk \, b_4^\sfk\,  z_\sfk (x,v),\qquad
 \Theta_4(x,v)=\sum_\sfk \, b_4^\sfk\, \vartheta_\sfk (x,v),
 \end{align}
where the modes $(z_\sfk ,\vartheta_\sfk )$ span a basis of solutions,
with $\sfk$ labeling different modes.  Different solutions are
parametrized by the expansion coefficients $b_4^\sfk $. Because the
equations for the pairs $(Z_1,\Theta_2)$ and $(Z_2,\Theta_1)$ have the
same form as the one for $(Z_4,\Theta_4)$, they must be expandable in
the same modes:
 \begin{align}
 Z_1    &=\sum_\sfk \, b_1^\sfk \,z_\sfk ,\qquad
 \Theta_2=\sum_\sfk \, b_1^\sfk \,\vartheta_\sfk ,\\
 Z_2    &=\sum_\sfk \, b_2^\sfk \,z_\sfk ,\qquad
 \Theta_1=\sum_\sfk \, b_2^\sfk \,\vartheta_\sfk .
 \end{align}
\end{subequations}

When the number of supergravitons in \eqref{gen_state_R_same_base} is
small, namely if $N^{\psi}_{k,m,n,f}\ll N$, we are in a linear regime
and $b_I^\sfk $ are also small; they are linearly related to
$(N^\psi_{k,m,n,f})^{1/2}$, or more precisely to the parameters
$A^\psi_{k,m,n,f}$ in the coherent state
\eqref{coh_superposition}.\footnote{Generally, the large $N$ scaling is
$b_I^\sfk\sim A^\psi_{k,m,n,f}N^{-1/2}\sim(N^{\psi}_{k,m,n,f}/N)^{1/2}$
\cite{Skenderis:2006ah, Kanitscheider:2006zf, Kanitscheider:2007wq}.
\label{ftnt:c_vs_A}} In the linear regime, the source terms on the
right-hand side of the second-layer equations \eqref{L2eqs} (or
\eqref{L2eqs'}) vanish, because they are quadratic in the first-layer
fields.  This means that the second-layer fields are the ones that
correspond to the base $\Psi_{\rm bg}$.

When the number of supergravitons in \eqref{gen_state_R_same_base} is
not small, namely if $N^{\psi}_{k,m,n,f}=\cO(N)$, the parameters
$b_I^\sfk $ are finite.  Because the first-layer equations are linear,
Eqs.~\eqref{L1_expn_general} are a valid solution even for finite
$b_I^\sfk $.  In this case, the second-layer equations have
non-vanishing source and the solution becomes non-trivial.  Also, the
linear relation between  $b_I^\sfk $ and $A^\psi_{k,m,n,f}$ gets
non-linear correction.  This correction must be in a very specific form
so that the full geometry is regular.  This is a powerful constraint
which in some cases can be used to determine the form of the non-linear
correction, without input from CFT\@.  We will see how this mechanism
(``coiffuring'') works in explicit examples below.

Because the source terms in the second-layer equations are quadratic in
the first-order fields, the expansion \eqref{L1_expn_general} means that
the source has the schematic form
\begin{align}
 \sum_{\sfk , \sfk '} \,b^\sfk \,b^{\sfk'}\text{(some function)},
\label{L2_src_expn_general}
\end{align}
where we ignored the structure related to the indices $I,I'$.
Therefore, we only have to solve the second-layer equations for each
pair of modes $(\sfk,\sfk')$; let us call the resulting second-layer fields
$\cF_{\sfk ,\sfk '}$, $\omega_{\sfk ,\sfk '}$.  Once we have them,
we can construct the solution for the general first-layer fields
\eqref{L1_expn_general} by superposing solutions for different pairs of
modes as $\sum_{\sfk ,\sfk '}b^\sfk b^{\sfk'}
\cF_{\sfk ,\sfk '}$.  The working assumption here is that, if
we can make the geometry regular for each pair of modes,  the
general solution obtained by superposition is also regular.  This does
work for known solutions \cite{Bena:2015bea}.

%

For the arguments above, the assumption that the base is unchanged, no
matter what $b_I^{\sfk }$ are, is crucial.  If that is not the case, we
will have to change the mode functions $z_{\sfk },\vartheta_{\sfk }$ as
we change $b_I^{\sfk }$, and the expansions \eqref{L1_expn_general} lose
their meaning as linear superposition.  In this article, we will
restrict ourselves to the cases where this assumption holds.  However,
we emphasize that this is a technical assumption that does not hold true
for the completely general superstrata dual to the CFT state
\eqref{1/8_sgrvtn_R}.  For constructing such general solutions, we would
have to face the problem of changing the base depending on the state.

\subsection{A class of superstrata with a flat base}
\label{ss:00_superstrata}

\medskip
\subsubsection{Linear spectrum}
\label{sss:linear_spectrum}

As discussed above, it is interesting to focus on 1/8-BPS excitations
that do not change the base.  To find the candidate states for which
that is true, let us look at the spectrum of linearized supergravity
around ${\rm AdS}_3\times S^3$. This means that we are taking the base
to be the flat base \eqref{base_metric_flat} equipped with
$\beta=\beta_0$ of \eqref{beta0} and the CFT background $\Psi_{\rm bg}$
to be \eqref{emptyAdS3_cft_state}.\footnote{If one wants to consider
some other background state $\Psi_{\rm bg}$, then one needs to study the
spectrum of linearized supergravity around the background geometry dual
to $\Psi_{\rm bg}$, in order to carry out the procedure of this section.
}

The spectrum of linearized supergravity around ${\rm AdS}_3\times S^3$ has
been long known \cite{Deger:1998nm, Maldacena:1998bw, Larsen:1998xm,
deBoer:1998kjm}.  As mentioned before, chiral primary states in CFT are
in one-to-one correspondence with 1/4-BPS supergraviton states in linearized
supergravity.  Likewise, $SU(1,1|2)_L\times SU(1,1|2)_R$ descendants of
chiral primary states are in one-to-one correspondence with 
1/8-BPS supergraviton states.

Let us see how these states are expressed in terms of the ansatz data of
section~\ref{sec:sugra_setup}.  The spectrum of 1/8-BPS supergravitons
and the non-trivial fields that they involve were worked out in
\cite{Deger:1998nm} in $d=6$ supergravity and reinterpreted in
\cite{Ceplak:2018pws} in the formulation of sections~\ref{sec:CFT} and~\ref{sec:sugra_setup}.  From \cite[Appendix C]{Ceplak:2018pws}, we see
that the following fields in the zeroth and first layers get excited:
\begin{subequations} 
 \begin{align}
 \ket{--'},\ket{\mp\pm},GG\ket{--'},GG\ket{-+}&:\quad ds^2(\cB),\beta,Z_I,\Theta_I\\
 \ket{00},\ket{++'},GG\ket{00},GG\ket{++'}  &:\quad  Z_I,\Theta_I\label{lin_spec_fixed_base}\\
 GG\ket{+-}&:\quad ds^2(\cB)
 \end{align}
\label{fuva23Jan20}
\end{subequations}
Here, $\ket{\psi}$ represents states of the form
$(L_{-1}-J_{-1}^3)^n(J_{-1}^+)^m\ket{\psi}_k$ and $GG\ket{\psi}$
represents states of the form
$(L_{-1}-J_{-1}^3)^n(J_{-1}^+)^mG_{-1}^{+,1}G_{-1}^{+,2}\ket{\psi}_k$,
in the R sector.  Also, $\ket{++'}$ is a particular superposition of
$\ket{++}$ and $\ket{--}$ which corresponds to the ``density mode'' of
the 2-charge solution \eqref{LM_geom_general} that changes the
$\lambda$-parametrization of the profile but not its shape \cite{Shigemori:2013lta, Bena:2016agb}.  On the other hand, $\ket{--'}$
is a superposition that is linearly independent of $\ket{++'}$, namely,
the ``transverse mode'' which changes the shape of the profile. We see
that, at the linear level, the states listed in
\eqref{lin_spec_fixed_base} do not change the base.  If we consider
non-linear correction, there is no guarantee that the base stays
undeformed. However, we will see that they in fact lead to superstrata
with a fixed base.

\subsubsection{A class of superstrata with a flat base}

Based on the linear spectrum \eqref{lin_spec_fixed_base}, let us
consider the class of superstrata that corresponds to the following set
of states:
\begin{subequations} 
 \label{circular_class_1}
 \begin{gather}
 [\ket{++}_1]^{N_0} 
 \prod_{k,m,n}\biggl\{
 \Bigl[\ket{00;k,m,n}\Bigr]^{N_{k,m,n}}
 \Bigl[\ket{00;k,m,n,12}\Bigr]^{\Nh_{k,m,n}}
 \biggr\},\\
 N_0+ \sum_{k,m,n} (k N_{k,m,n} + k \Nh_{k,m,n}) = N.
 \end{gather}
\end{subequations}
Here we wrote $N^{00}_{k,m,n}=:N_{k,m,n}$,
$N^{00}_{k,m,n,12}=:\Nh_{k,m,n}$.  The first factor $[\ket{++}_1]^{N_0}$
is the background part $\Psi_{\rm bg}$ in \eqref{gen_state_R_same_base},
and we wrote its total strand length as $N_0:= N_{\rm bg}$.  This
corresponds to empty ${\rm AdS}_3\times S^3$, as in
\eqref{emptyAdS3_cft_state}.
In this case, although there is no general construction or proof yet,
experience shows \cite{Bena:2015bea, Bena:2016agb, Bena:2016ypk,
Bena:2017geu, Bena:2017upb, Bena:2017xbt, Ceplak:2018pws,
Heidmann:2019zws}
that we can take the base $\cB$ to be
flat $\bbR^4$ with metric \eqref{base_metric_flat} and the 1-form
$\beta$ to be $\beta_0$ given in \eqref{beta0}.
In this subsection, we will discuss the superstratum solutions dual to
the class of state  \eqref{circular_class_1}.
Although this is a subclass of all possible superstrata, their entropy
growth rate for large charges is expected to be the same as that for
more general superstrata ensemble, $S\sim N^{1/2}N_P^{1/4}$
\cite{Shigemori:2019orj}.

In \eqref{lin_spec_fixed_base}, we also have states based on
$\ket{++'},GG\ket{++'}$.  They will be discussed in section~\ref{sss:strata_other_species}.

\subsubsection{Linear solutions}

If the excitation numbers $N_{k,m,n},\Nh_{k,m,n}$ in the state
\eqref{circular_class_1} are much smaller than $N$, it describes small
fluctuations around ${\rm AdS}_3\times S^3$; namely, solutions of
linearized supergravity in the ${\rm AdS}_3\times S^3$ background. The
explicit form of such linear solutions in $d=6$ supergravity is known
\cite{Deger:1998nm, Larsen:1998xm} and we can read off the ansatz data
from them, although that requires knowledge of how our ansatz is
embedded in $d=6$ supergravity.

Another way to find the ansatz data for linear solutions is the
so-called \emph{solution-generating technique}~\cite{Mathur:2003hj},
which was used in constructing explicit superstrata \cite{Bena:2015bea,
Bena:2017xbt, Ceplak:2018pws}.  The AdS/CFT dictionary for 1/4-BPS
supergravitons is given in \eqref{holog_dict_2-chg}, so we know the
linear solution dual to $\ket{\psi}_k$.  For example, the dual of
$\ket{00}_k$ is obtained from \eqref{Z4_excitaion_ansatz_data} by taking
infinitesimal~$b$.  By the bulk spectral flow \eqref{bulk_spectr_flow},
we can transform the background to ${\rm AdS}_3\times S^3$ with the
super-isometry group $SU(1,1|2)_L\times SU(1,1|2)_R$, which is dual to
the symmetry group of the boundary CFT\@.  Its bosonic generators,
including $L_{-1},J_0^{-}$,\footnote{Here we using the NS language,
appropriate for the ${\rm AdS}_3\times S^3$ background.} can be realized
in the bulk as Killing vectors and, by acting with the corresponding
diffeomorphism on the linear solution dual to $\ket{00}_k$, we can
obtain the linear solution dual to $(L_{-1})^n
(J_0^-)^{m}\ket{00}_k=\ket{00;k,m,n}$.  If we reorganize the resulting
linear solution in the form of the ansatz \eqref{ansatzSummary}, we can
read off the ansatz data \cite{Shigemori:2013lta, Bena:2015bea,
Bena:2017xbt}.  Likewise, the fermionic generators $G_{-1/2}^{-,A}$ are
realized as the supersymmetry transformations with Killing spinors
preserved by the ${\rm AdS}_3\times S^3$ background, and its action
allows us to construct the linear solution dual to $(L_{-1})^n
(J_0^-)^{m}G_{-1/2}^{-,1}G_{-1/2}^{-,2}\ket{00}_k=\ket{00;k,m,n,12}$ and
read off the corresponding ansatz data~\cite{Ceplak:2018pws}.

This procedure leads to the following ansatz data:
\begin{subequations} 
\label{linear_sol_Z4Theta_4}
 \begin{align}
 &\ket{00;k,m,n}&
 &\hspace*{-10ex}\leftrightarrow \qquad
 Z_4=b_4 z_{k,m,n},\quad
 \Theta_4= b_4 \vartheta_{k,m,n},\\
 &\ket{00;k,m,n,12}&
 &\hspace*{-10ex}\leftrightarrow \qquad
 Z_4=0,\qquad~~~~~
 \Theta_4= \bh_4 \varthetah_{k,m,n},
 \end{align}
\end{subequations}
where $b_4,\bh_4$ are small constants which are proportional to
$(N_{k,m,n})^{1/2},(\Nh_{k,m,n})^{1/2}$, or more precisely the
parameters $A_{k,m,n}\equiv A^{00}_{k,m,n},\Ah_{k,m,n}\equiv
A^{00}_{k,m,n,12}$ that appear in the coherent superposition in
\eqref{coh_superposition} (see footnote \ref{ftnt:c_vs_A}).  All other
fields, $ds^2(\cB), \beta, Z_{1,2},\Theta_{1,2},\omega,\cF$ are still
given by the empty ${\rm AdS}_3\times S^3$ ones,
\eqref{ansatz_data_empty_AdS3xS3}. In particular, the base is
undeformed. The explicit form of the mode function $z_{k,m,n}$ and
2-forms $\vartheta_{k,m,n},\varthetah_{k,m,n}$ are given by
\begin{subequations} 
 \label{z_theta_kmn}
 \begin{align}
 z_{k,m,n} 
 &\equiv R_y \frac{\Delta_{k,m,n}}{\Sigma}\cos{v_{k,m,n}},
  \label{z_kmn}
\\
 \vartheta_{k,m,n}&\equiv -\sqrt{2}\,
 \Delta_{k,m,n}
 \biggl[\biggl((m+n)\,r\sin\theta +n\left({m\over k}-1\right){\Sigma\over r \sin\theta}  \biggr)\Omega^{(1)}\sin{v_{k,m,n}} \nonumber\\
 &\hspace{16ex}
 +\biggl(m\left({n\over k}+1\right)\Omega^{(2)} +\left({m\over k}-1\right)n\, \Omega^{(3)}\biggr) \cos{v_{k,m,n}} \biggr]
    \label{theta_kmn}
 \,,
  \\
   \varthetah_{k,m,n} &\equiv
            \sqrt{2}  \Delta_{k, m, n}\left[\frac{\Sigma}{r\sin\theta}\Omega^{(1)} \sin{\hat v_{k,m,n}}+ \left(\Omega^{(2)} + \Omega^{(3)}\right)\cos{\hat v_{k,m,n}}\right]\,.
    \label{thetah_kmn}
 \end{align}
\end{subequations}
where
\begin{subequations} 
 \label{Delta_v_kmn_def}
 \begin{align}
  \Delta_{k,m,n} &\equiv
 \left(\frac{a}{\sqrt{r^2+a^2}}\right)^k
 \left(\frac{r}{\sqrt{r^2+a^2}}\right)^n 
 \cos^{m}\theta \, \sin^{k-m}\theta , 
 \label{Delta_kmn_def}
 \\
 v_{k,m,n} &\equiv (m+n) \frac{\sqrt{2}\,v}{R_y} + (k-m)\phi - m\psi ,
 \label{v_kmn_def}
 \end{align}
\end{subequations}
and the $\Omega^{(i)}$ are a (unnormalized) basis of self-dual
$2$-forms:
\begin{equation}
\label{selfdualbasis}
\begin{aligned}
\Omega^{(1)} &\equiv \frac{dr\wedge d\theta}{(r^2+a^2)\cos\theta} + \frac{r\sin\theta}{\Sigma} d\phi\wedge d\psi\,,\\
\Omega^{(2)} &\equiv  \frac{r}{r^2+a^2} dr\wedge d\psi + \tan\theta\, d\theta\wedge d\phi\,,\\
 \Omega^{(3)}&\equiv \frac{dr\wedge d\phi}{r} - \cot\theta\, d\theta\wedge d\psi\,.
\end{aligned}
\end{equation}
One can check that \eqref{linear_sol_Z4Theta_4} satisfy the first-layer
equations \eqref{L1eqs-4'} for the flat base \eqref{base_metric_flat}.
More generally, one could include constant phase in \eqref{z_theta_kmn}
by setting $v_{k,m,n}\to v_{k,m,n}+\alpha_{k,m,n}$, but we do not
consider such generalization here.

Because we are in linearized supergravity, we can freely take a linear
superposition of different modes in \eqref{linear_sol_Z4Theta_4},
obtaining
\begin{align}
 Z_4=\sum_{k,m,n}b_4^{k,m,n} z_{k,m,n},\qquad
 \Theta_4= \sum_{k,m,n}\left(
 b_4^{k,m,n} \vartheta_{k,m,n}+
\bh_4^{k,m,n} \varthetah_{k,m,n}\right).
\label{Z4Th4_small_b4}
\end{align}
where $b_4^{k,m,n}$ and $\bh_4^{k,m,n}$ are infinitesimal and
proportional to $A_{k,m,n}$ and $\Ah_{k,m,n}$, respectively.  All other
fields remain undeformed.  This is the solution that corresponds to the
state \eqref{circular_class_1} with general $N_{k,m,n},\Nh_{k,m,n}\ll
N$.

\subsubsection{Non-linear solutions and coiffuring}

If the excitation numbers $N_{k,m,n},\Nh_{k,m,n}$ are $\cO(N)$ (or
$A_{k,m,n},\Ah_{k,m,n}=\cO(N^{1/2})$), we must go beyond the linear
approximation and consider backreaction.  Such non-linear solutions are
nothing but superstrata.  We must find a solution to all three
layers of BPS equations, generalizing the linear solution
\eqref{Z4Th4_small_b4}.  The working assumption in doing so is that,
even in such a non-linear regime, the base remains undeformed and is
still given by~\eqref{base_metric_flat} and~\eqref{beta0}.
Because the first-layer equations are linear, we can still use
\eqref{Z4Th4_small_b4} but now with the coefficient
$b_4^{k,m,n},\bh_4^{k,m,n}$ \emph{finite}.  This means that the source
terms will be non-vanishing in the second-layer equations, which we must
solve, imposing regularity.

At linear order, $b_4^{k,m,n},\bh_4^{k,m,n}$ were proportional to
$A^{k,m,n},\Ah^{k,m,n}$.  However, in the non-linear regime, there can
be non-linear corrections to the relation.  Moreover, there can be
non-linear correction to other first-layer fields,
$Z_{1,2},\Theta_{1,2}$.  Because $(Z_1,\Theta_2)$, $(Z_2,\Theta_1)$
satisfy the same equation satisfied by $(Z_4,\Theta_4)$, we must be able
to expand $Z_I,\Theta_I$ as
\begin{subequations} 
 \begin{align}
 Z_1&={R_y^2a^2\over Q_5\Sigma}+\sum_{k,m,n}b_1^{k,m,n} z_{k,m,n},&
 \Theta_2&=
 \sum_{k,m,n}(b_1^{k,m,n} \vartheta_{k,m,n}+\bh_1^{k,m,n} \varthetah_{k,m,n}),
\label{Z1Th2_gen_expn}
 \\
 Z_2&={Q_5\over \Sigma}+\sum_{k,m,n}b_2^{k,m,n} z_{k,m,n},&
 \Theta_1&=
 \sum_{k,m,n}(b_2^{k,m,n} \vartheta_{k,m,n}+\bh_2^{k,m,n} \varthetah_{k,m,n}),
\label{Z2Th1_gen_expn}
 \\
 Z_4&=\sum_{k,m,n}b_4^{k,m,n} z_{k,m,n},&
 \Theta_4&=
 \sum_{k,m,n}(b_4^{k,m,n} \vartheta_{k,m,n}+\bh_4^{k,m,n} \varthetah_{k,m,n}),\label{Z4Th4_gen_expn}
 \end{align}
\end{subequations}
where in $Z_{1,2}$ we included ``zero mode'' terms from
\eqref{emptyAdS3_Z_I}.  $b_I,\bh_I$ are finite numbers with
$b_4=\cO(A),\bh_4=\cO(\Ah)$ and
$b_1,\bh_1,b_2,\bh_2=\cO(A^2,\Ah^2,A\Ah)$. Or, alternatively, we can
write the relation as $b_1,\bh_1,b_2,\bh_2=\cO(b_4^2,\bh_4^2,b_4\bh_4)$.
These non-linear corrections in the first layer feed into the second
layer as source.  We must solve the second-layer equations with the
source, and determine the correction so that the full solution represent
a regular geometry.  In principle, the coefficients can receive
corrections from all orders:
\begin{align}
 B_1^{\bm{k}}
 &\stackrel{?}{=}
 \sum_{\bm{k}_1,\bm{k}_2}
 c^{\bm{k}}_{\bm{k}_1,\bm{k}_2}
 B_4^{\bm{k}_1}B_4^{\bm{k}_2}
 +
 \sum_{\bm{k}_1,\bm{k}_2,\bm{k}_3}
 c^{\bm{k}}_{\bm{k}_1,\bm{k}_2,\bm{k}_3}
 B_4^{\bm{k}_1}B_4^{\bm{k}_2}B_4^{\bm{k}_3}
+\cdots
\label{b1_corr_b4_inpr}
\end{align}
where ${\bm k}=(k,m,n)$ and $B_I^{\bm{k}}$ collectively denotes
$b_I^{\bm{k}}$ and $\bh_I^{\bm{k}}$. Namely, even if one turns on
$B_4^{\bm{k}}$ for one or two particular values of $\bm{k}$, it can make
$B_1^{\bm{k}},B_2^{\bm{k}}$ non-vanishing for infinitely many values of
$\bm{k}$. Also, it is expected that the corrections are not unique, due
to the possibility to turn on new states at higher order.  Determining
all the corrections seems to be a formidable task.

However, fortunately, we can gain an idea about how to proceed by using
a finite version of the solution-generating technique
\cite{Giusto:2013bda, Bena:2015bea}.  Namely, one starts with the
Lunin-Mathur geometry with profile \eqref{circularLM+g5} with finite
$b$, and furthermore acts on it with a finite $SU(2)_L$ rotation.  This
procedure generates a particular solution of all three layers and gives
us an idea about what the general solution must look like.  From this,
we can extract a rule of thumb, called \emph{coiffuring}
\cite{Bena:2015bea} , which can be stated as follows.  Let
$(Z_4,\Theta_4)$ be given by \eqref{Z4Th4_gen_expn} with general finite
coefficients $B_4^{\bm{k}}=(b_4^{\bm{k}},\bh_4^{\bm{k}})$.  First, we
set $B_2^{\bm{k}}=(b_2^{\bm{k}},\bh_2^{\bm{k}})=0$. Then, we choose
$B_1^{\bm{k}}=(b_1^{\bm{k}},\bh_1^{\bm{k}})$ as follows.  If
$(Z_4,\Theta_4)$ have modes $\bm{k}_1=(k_1,m_1,n_1)$ and
$\bm{k}_2=(k_2,m_2,n_2)$ turned on, they will produce, when fed into the
second-layer equations \eqref{L2eqs'}, sources with ``high-frequency''
phase $v_{\bm{k}_1+\bm{k}_2}\equiv v_+$ and ``low-frequency'' phase
$v_{\bm{k}_1-\bm{k}_2}\equiv v_-$.  This is because the source in
\eqref{L2eqs'} includes quadratic terms in $Z_4,\Theta_4$, and because
of the product formula for trigonometric functions.  Because a
high-frequency source leads to a singularity in the 1-form~$\omega$, we
must set the coefficients in $(Z_1,\Theta_2)$ so that the high-frequency
terms get canceled in the source. Because $Z_2$ has a zero-mode term
($Q_5/\Sigma$), this can be achieved by setting
$B_1^{\bm{k}_1+\bm{k}_2}$ in $(Z_1,\Theta_2)$ to be proportional to
$B_4^{\bm{k}_1}B_4^{\bm{k}_2}$.
The actual procedure of coiffuring can be messy and the detail depends
on the values of the mode numbers $\bm{k}_1,\bm{k}_2$.  If
$\bm{k}_1-\bm{k}_2$ is an allowed wave number, we may also have to turn
on $B_1^{\bm{k}_1-\bm{k}_2}\propto B_4^{\bm{k}_1}B_4^{\bm{k}_2}$ in
order to cancel the low-frequency source with phase $v_-$, to avoid a
singular term in~$\omega$.\footnote{This coiffuring for low-frequency
source is more non-trivial than the high-frequency one.  For
low-frequency coiffuring, the term in $Z_1$ to be turned on is
proportional to $\Delta_{k_1-k_2,m_1-m_2,n_1-n_2}$, whereas one naively
expects terms proportional to $\Delta_{k_1,m_1,n_1}\Delta_{k_2,m_2,n_2}=
\Delta_{k_1+k_2,m_1+m_2,n_1+n_2}$, the second-layer source being
quadratic in $Z_I,\Theta_I$.}
In any case,~$B_1^{\bm{k}}$ are quadratic in
$B_4^{\bm{k}}$, and the expansion \eqref{b1_corr_b4_inpr} actually
terminates at quadratic order.  Also, if one turns on $B_4^{\bm{k}}$ for
the pair $(\bm{k}_1,\bm{k}_2)$, it makes $B_1^{\bm{k}}$ non-vanishing
only for finite (actually, up to two) values of $\bm{k}$.  Therefore, as
mentioned below \eqref{L2_src_expn_general}, we only have to solve the
second-layer equations for each pair of modes $(\bm{k}_1, \bm{k}_2)$ to
find a regular solution for general $(Z_4,\Theta_4)$ in
\eqref{Z4Th4_gen_expn}.  If we can find the solution to the second
layer, call it $\cF_{\bm{k}_1,\bm{k}_2},\omega_{\bm{k}_1,\bm{k}_2}$, for
the pair $(\bm{k}_1, \bm{k}_2)$, the solution $\cF,\omega$ for the
general case \eqref{Z4Th4_gen_expn} can be obtained by summing over all
pairs.

At the time or writing, no closed formula for the coiffured $Z_1$ for a
general pair of modes $(\bm{k}_1, \bm{k}_2)$ and the resulting
second-layer fields $\omega,\cF$ is known.  However, for some sets of pairs of modes
(some of which are infinite sets), coiffuring has been explicitly
carried out and the full solution has been shown to be completely
regular.  The interested reader are referred to \cite{Bena:2015bea,
Heidmann:2019zws} for detail.

Precision holography \cite{Giusto:2015dfa, Giusto:2019qig} indicates
that, the relation between the mode coefficients
$b_4^{k,m,n},\bh_4^{k,m,n}$ and the coherent state parameters
$A_{k,m,n},\Ah_{k,m,n}$ is not modified at higher order; they are simply
proportional to each other.  This suggests that coiffuring is the way
preferred by CFT to fix the mode coefficients.  This is presumably
related to the fact that coiffuring is in some sense the minimal way to
achieve regularity by writing $B_1^{\bm{k}}$ as a mere quadratic
expression in $B_4^{\bm{k}}$.

In the early stages of the development, attempts were made to construct
smooth superstrata based on states that change the shape of the
Lunin-Mathur geometries, $\ket{\alpha\dot\alpha}_k$ where
$\alpha,\dot\alpha=\pm$, but only singular solutions were obtained
\cite{Niehoff:2013kia}.  In retrospect, having a fixed base and $\beta$
is technically much easier and, turning on $g_5$ dual to $\ket{00}_k$
is the most natural way to go.  However, even so, it is miraculous that
coiffuring allows us to explicitly construct a class of superstrata for
which the base is fixed, no matter what modes $(k,m,n)$ we turn on.
This could have failed at any stage, because it could be that, once the
deformation is finite, regularity requirement inevitably leads to
uncontrollable non-linear correction to the first-layer fields, or even
to deformation of the base.  Currently we lack a deep understanding of
why coiffuring works.

\subsection{Explicit superstratum solutions}
\label{ss:explicit_strata}

\medskip

\subsubsection{Single-mode superstrata}

In section \ref{ss:00_superstrata}, we explained how to construct
superstrata dual to the CFT states of the form~\eqref{circular_class_1}
for general $N_{k,m,n}, \Nh_{k,m,n}$.  By coiffuring, the construction
reduces to solving the BPS equations for a general pair of modes
$(k_1,m_1,n_1)$ and $(k_2,m_2,n_2)$ in $Z_4$.  For the solutions of the
class \eqref{circular_class_1}, such multi-mode superstrata have been
constructed for some particular pair of modes on a case-by-case basis,
but the general solution has not been found yet at the time of writing
(see \cite{Heidmann:2019zws} for recent development).  So, just as in
much of the literature, we will mostly focus on \emph{single-mode}
superstrata, for which only one particular mode $(k,m,n)$ is turned on.
We will mention the multi-mode case when appropriate.

So, we take the first-layer fields to be
\begin{align}
  Z_4&=b_4\, z_{k,m,n},\qquad
 \Theta_4= b_4\, \vartheta_{k,m,n}+\bh_4\, \varthetah_{k,m,n},
\end{align}
where we have suppressed the mode index on the coefficients; more
precisely $b_4,\bh_4$ must be written as $b_4^{k,m,n},\bh_4^{k,m,n}$.
Coiffuring is not trivial even in this case, because the mode $(k,m,n)$
quadratically interacts with itself.
The CFT state dual to this superstratum is the same as~\eqref{circular_class_1} but without the product.  Namely,
\begin{subequations} 
 \begin{gather}
 [\ket{++}_1]^{N_0} 
 \Bigl[\ket{00;k,m,n}\Bigr]^{N_{k,m,n}}
 \Bigl[\ket{00;k,m,n,12}\Bigr]^{\Nh_{k,m,n}},
  \label{gjpo11Jan20}
\\
 N_0+ k (N_{k,m,n} + \Nh_{k,m,n}) = N.\label{irns7Jan20}
 \end{gather}
\end{subequations}

The ``original'' superstratum constructed in
\cite{Bena:2015bea,Bena:2016ypk, Bena:2017xbt} corresponds to the one
with $b_4\neq 0$, $\bh_4=0$, while the ``supercharged''
superstratum constructed in \cite{Ceplak:2018pws} corresponds to the one
with $b_4=0$, $\bh_4\neq 0$.  The case with
$b_4,\bh_4\neq 0$ is called the ``hybrid'' superstratum
and was constructed in \cite{Heidmann:2019zws}.  Our presentation of the
solution follows \cite{Heidmann:2019zws}.

\subsubsection{The first layer}

In the presence of a single mode $\bm{k}=(k,m,n)$, the second-layer
source will have high-frequency terms with phase
$v_{2\bm{k}}=2v_{\bm{k}}$ and low-frequency terms with constant phase.
The constant-phase terms, or the ``RMS'' terms, do not lead to
singularities in $\omega,\cF$, while the high-frequency terms do and
must be coiffured away.  This means that we must turn on a term with
mode numbers~$2\bm{k}$ in $Z_1$.  Therefore, we are led to the following
ansatz for the first-layer fields:
\begin{align}
 \label{eq:hyb_strata_L1data}
 \begin{aligned}
 Z_1 &= \frac{R_y^2 a^2}{Q_5\Sigma} + b_1 \frac{R_y }{2Q_5}  z_{2 k,2m,2n},
  &
  \Theta_2  &=  
  b_1 \frac{R_y}{2 Q_5} \vartheta_{2k,2m,2n}
  +\bh_1 \frac{R_y}{2 Q_5} \varthetah_{2k,2m,2n}
  , \\
  Z_2 &=\frac{Q_5}{\Sigma} ,  &
  \Theta_1  &=  0, \\
   Z_4  &=  b_4 z_{k,m,n},&
   \Theta_4 &=   b_4 \vartheta_{k,m,n}+\bh_4 \varthetah_{k,m,n},
 \end{aligned}
\end{align}
where the coefficients  $b_1,\bh_1$ in $Z_4$ must more properly be written as
$b_1^{2k,2m,2n}$, $\bh_1^{\,2k,2m,2n}$.

\subsubsection{The second layer}

If we plug in the ansatz \eqref{eq:hyb_strata_L1data} into the
second-layer equations \eqref{L2eqs'}, we find that the high-frequency
terms cancel if we the coefficients satisfy
 the following coiffuring constraints:
\begin{align}
 b_1=b_4^2,\qquad \bh_1=2b_4\bh_4.\label{coiff_single-mode_hyb}
\end{align}

Let us solve the second-layer equations.
When the coiffuring relation \eqref{coiff_single-mode_hyb} is
satisfied, the source in the second-layer equation \eqref{L2eqs'}
consists only of an RMS term.  So, $\omega$ and $\cF$ can be written as
\begin{align} 
\omega = \omega_0  + \omega_{k,m,n},  \qquad  
 \cF = \cF_{k,m,n}
\end{align} 
where $\omega_{k,m,n},\cF_{k,m,n}$ are RMS modes, independent of
$v_{k,m,n}$.  The second-layer equations \eqref{L2eqs'} are
\begin{align}
& (1+*_4)d \omega_{k,m,n} + \mathcal{F}_{k,m,n} \,d\beta
 =
 \sqrt{2}R_y \frac{\Delta_{2k,2m,2n}}{\Sigma} 
 \biggl[
 \biggl(\frac{m(k+n)}{k} b_4-\bh_4\biggr)\Omega^{(2)} 
 \notag\\
 &\hspace*{50ex}
 - \biggl(\frac{n(k-m)}{k} b_4+\bh_4\biggr)\Omega^{(3)} 
 \biggr] , 
 \label{eq:omegakmn}\\
 \label{eq:Fkmn}
 &\widehat{\mathcal{L}} \,  \cF_{k,m,n}
=
 \frac{4}{(r^2+a^2)\Sigma \cos^2\theta } 
 \biggl[ 
   \biggl(\frac{m(k+n)}{k} b_4-\bh_4\biggr)^2\Delta_{2k,2m,2n} \notag\\
 &\hspace*{30ex}
 + \biggl(\frac{n(k-m)}{k} b_4+\bh_4\biggr)^2 \Delta_{2k,2m+2,2n-2} 
 \biggr],
\end{align}
where $\widehat{\mathcal{L}}$ is the scalar Laplacian on the base 
$\cB=\bbR^4$:
\begin{equation}
\widehat{\mathcal{L}} F \equiv \frac{1}{r\Sigma}\, \partial_r \big( r (r^2 + a^2) \, \partial_r F  \big)  +    \frac{1}{\Sigma\sin \theta \cos \theta}\partial_\theta \big( \sin \theta \cos \theta\, \partial_\theta F  \big)\,.
\end{equation}

The solution to Eq.~\eqref{eq:Fkmn} is given by
\begin{align}
 \cF_{k,m,n} &= 4
 \biggl[
  \biggl(\frac{m(k+n)}{k}b_4-\bh_4\biggr)^2F_{2k,2m,2n}
 +\biggl(\frac{n(k-m)}{k}b_4+\bh_4\biggr)^2F_{2k,2m+2,2n-2}
 \biggr],
\end{align}
where $F_{2k,2m,2n}$ solves the equation
\begin{align}
 \widehat{\cL}F_{2k,2m,2n}={\Delta_{2k,2m,2n}\over (r^2+a^2)\cos^2\theta\,\, \Sigma}
\end{align}
and its explicit form is \cite{Bena:2017xbt}
\begin{align} 
F_{2k,2m,2n}\,=\,-\!\sum^{j_1+j_2+j_3\le k+n-1}_{j_1,j_2,j_3=0}\!\!{j_1+j_2+j_3 \choose j_1,j_2,j_3}\frac{{k+n-j_1-j_2-j_3-1 \choose k-m-j_1,m-j_2-1,n-j_3}^2}{{k+n-1 \choose k-m,m-1,n}^2}\,
\frac{\Delta_{2(k-j_1-j_2-1),2(m-j_2-1),2(n-j_3)}}{4(k+n)^2(r^2+a^2)}\label{def_F_2k,2m,2n}
\end{align} 
with  the multinomial coefficients defined by
\begin{align} 
{j_1+j_2+j_3 \choose j_1,j_2,j_3}\equiv \frac{(j_1+j_2+j_3)!}{j_1!\, j_2!\, j_3!}.
\end{align} 

On the other hand, $\omega_{k,m,n}$ can be written as
\begin{align}
\omega_{k,m,n} =  \mu_{k,m,n} (d\psi+d\phi) + \zeta_{k,m,n}(d\psi-d\phi)\,.
\label{eq:omkmnparts1}
\end{align}
If we define
\begin{align}
\muh_{k,m,n} \equiv \mu_{k,m,n} +\frac{R_y}{4\sqrt{2}}\frac{r^2+a^2\sin^2\theta}{\Sigma}\mathcal{F}_{k,m,n}+\frac{R_y\,b_4^2}{4\sqrt{2}} \,\frac{\Delta_{2k,2m,2n}}{\Sigma},
\end{align}
then $\muh_{k,m,n}$ is found to satisfy
\begin{align}
\widehat{\mathcal{L}}\, \muh_{k,m,n} &= 
 \frac{R_y}{\sqrt{2}}\frac{1}{(r^2+a^2) \Sigma\cos^2\theta} 
\biggl[
 \biggl(\frac{(k-m)(k+n)}{k}b_4+\bh_4\biggr)^2 \Delta_{2k,2m+2,2n}
 \notag\\
&\hspace*{25ex}
 +\biggl(\frac{m n}{k}b_4-\bh_4\biggr)^2 \Delta_{2k,2m,2n-2}
\biggr] .
\label{poisson_eq_mu-hat}
\end{align} 
Therefore,
\begin{align}
\mu_{k,m,n}&= \frac{R_y}{\sqrt{2}}\,\biggl[ 
\biggl(\frac{(k-m)(k+n)}{k}b_4+\bh_4\biggr)^2 F_{2k,2m+2,2n}
+\biggl(\frac{m n}{k}b_4-\bh_4\biggr)^2 F_{2k,2m,2n-2}
\notag\\
&\hspace{20ex}
-{b_4^2}\frac{\Delta_{2k,2m,2n}}{4\Sigma}\biggr]
 -{R_y\over 4\sqrt{2}} \frac{r^2+a^2\,\sin^2\theta}{\Sigma}\mathcal{F}_{k,m,n}
+\frac{R_y X_{k,m,n}}{2\sqrt{2}\,\Sigma}.
\label{mu}
\end{align} 
In the expression for $\mathcal{F}_{k,m,n}$ and $\mu_{k,m,n}$, it should
be understood that, when the coefficient of the $F$ function in a term
is zero, that term is zero; this rule is necessary because
$F_{2k,2m,2n}$ defined in \eqref{def_F_2k,2m,2n} can be ill-defined for
some values of $k,m,n$.  In $\muh$, the term proportional to
$X_{k,m,n}$ is a harmonic piece that can be freely added to the solution
of the Poisson equation~\eqref{poisson_eq_mu-hat}. 

Finally, once $\cF_{k,m,n}$ and $\mu_{k,m,n}$ are known, $\zeta_{k,m,n}$
is determined by the equations
\begin{align}
\begin{split}
 \partial_r \zeta_{k,m,n} &= 
 \frac{r^2 c_{2\theta}-a^2 s_{\theta}^2}{\Lambda} \partial_r\mu_{k,m,n}
 -\frac{r s_{2\theta}}{\Lambda}\partial_\theta \mu_{k,m,n}\\
 &\quad+\frac{\sqrt{2}R_y r}{\Sigma\Lambda}
 \biggl[ b_4\biggl( (m s^2_{\theta} + n c^2_{\theta})b_4 -\Bigl({m n\over k}b_4-\bh_4\Bigr) c_{2\theta} \biggr) \Delta_{2k,2m,2n}
 \\
 &\hspace*{20ex}
 -\frac{a^2 (2r^2+a^2)s^2_{\theta}c^2_{\theta}}{\Sigma}\mathcal{F}_{k,m,n}\biggr] ,\\
 \partial_\theta\zeta_{k,m,n}&=\frac{r(r^2+a^2) s_{2\theta}}{\Lambda}\partial_r \mu_{k,m,n}+ \frac{r^2 c_{2\theta}-a^2 s_{\theta}^2}{\Lambda} \partial_\theta\mu_{k,m,n}\\
 &\quad+\frac{R_y s_{2 \theta}}{\sqrt{2}\,\Sigma\Lambda}
 \biggl[ b_4 \biggl( (- m r^2+ n (r^2+a^2))b_4 - (2r^2+a^2)\Bigl({m n\over k}b_4-\bh_4\Bigr) \biggr) \Delta_{2k,2m,2n} \\
 &\hspace*{20ex}
  +\frac{a^2 r^2 (r^2+a^2)c_{2\theta}}{\Sigma}\mathcal{F}_{k,m,n}\biggr] ,
\end{split} 
\label{eq:zeta-eqns}
\end{align}
where $\Lambda\equiv r^2+a^2 \sin^2{\theta}$,
$s_\theta\equiv\sin\theta$, $c_\theta\equiv\cos\theta$.  For specific
values of $(k,m,n)$, it is easy to solve these to find explicit
$\zeta_{k,m,n}$, but the expression for $\zeta_{k,m,n}$ for general
$(k,m,n)$ is not known in closed form.

One obstacle to obtaining a general expression for $\zeta$ for general
$(k,m,n)$ is that the $F$ function is defined only through a sum in
\eqref{def_F_2k,2m,2n}.\footnote{The expression \eqref{def_F_2k,2m,2n}
can be regarded as a sort of triple hypergeometric function.  } If we
set two of the three variables $(k,m,n)$ to specific values, for example
as $(k,m,n)=(2,1,n)$, then the summation for $k,m$ is a finite sum while
the summation over $n$ is elementary to carry out; we can plug the
result into \eqref{eq:zeta-eqns} and find~$\zeta$.  However, for general
$(k,m,n)$, it is hard to evaluate the sum.

\subsubsection{Regularity}

For the solution to represent a microstate geometry, it must be regular
everywhere.  Fully establishing regularity, including absence of
closed-timelike curves, in full generality is quite challenging and only
a case-by-case results are known.  Here we discuss some salient features
important for the physics of the solutions.  Some more detail can be
found in, {\it e.g.}, \cite{Bena:2017xbt}.

First, the point $r=\theta=0$ is the origin of the flat $\bbR^4$ base
and all Cartesian components of all forms must be finite there.  In
particular, for the 1-form $\omega$ to be regular there, it is necessary
that its component along $d\phi+d\psi$, namely $\mu_{k,m,n}$ vanish
there.  This fixes the undetermined coefficient $X_{k,m,n}$ to be
\begin{align}
 \label{eq:X}
 X_{k,m,n}=B_4^2+\Bh_4^2,
\end{align}
where we defined $B_4,\Bh_4$ by
\begin{align}
 B_4^2={k+n\over 2k}
{k+n\choose k-m,\,m,\,n}^{-1}b_4^2,\qquad
 \Bh_4^2={k\over 2(k-m)m n} 
{k+n\choose k-m,\,m,\,n}^{-1}\bh_4^2
.
\end{align}
The component of $\omega$ along $d\phi+d\psi$, namely $\zeta_{k,m,n}$
must also vanish. This can be checked on a case-by-case basis for
values of $(k,m,n)$ for which explicit solutions and does not
lead to a new constraint.

Another dangerous point is $r=0,\theta=\pi/2$, where the original
2-charge supertube sits.    By requiring that the
$(d\phi+d\psi)^2$ component of the metric remain finite, one finds that
\begin{align}
 a^2+B_4^2+\Bh_4^2= {Q_1Q_5\over R_y^2}.
\label{gjtg11Jan20}
\end{align}
This can be thought of as the bulk version of the strand-budget
relation, \eqref{irns7Jan20}.  
The fact that one cannot excite gravity modes by an arbitrary amount is
sometimes called the stringy exclusion principle
\cite{Maldacena:1998bw}.  In linearized supergravity such constraint is
not visible, but in fully backreacted geometries it is known that such
constraint can arise by requiring the solution to be physical
\cite{deBoer:2009un}.

\subsubsection{Examples}
\label{sss:examples}

Although the explicit expression for single-mode superstrata with
general $(k,m,n)$ has not been found (because the expression for $\zeta$
is not known), some infinite families of solutions have been explicitly
written down in the literature.

For example, for $(k,m,n)=(1,0,n)$ \cite{Bena:2016ypk, Bena:2017xbt},
\begin{align}
\cF_{1,0,n} &= - \frac{b_4^2}{a^2} \, \bigg(1 - \frac{r^{2n}}{(r^2+a^2)^n}\bigg),\qquad
\omega_{1,0,n} =
 \frac{b_4^2 \, R_y}{\sqrt{2} \, \Sigma}\, \bigg(1 - \frac{r^{2n}}{(r^2+a^2)^n}\bigg) \, \sin^2\theta\, d\phi\,. 
\end{align}
In this case, there is no supercharged mode.  This solution is
deceivingly simple but has a quite non-trivial structure and contains
rich physics.  For $a\ll b_4$, this geometry is roughly AdS$_3$ for
$r\gtrsim b_4$.  For $a\lesssim r\lesssim b_4$, the spacetime is an
AdS$_2\times S^1$ throat.  At $r\sim a$, there is a momentum wave that
supports the geometry which smoothly caps off at $r=0$.  For studies of
various physical aspects of this solution, see \cite{Bena:2016ypk,
Bena:2017upb, Tyukov:2017uig, Bena:2017xbt, Raju:2018xue, Bena:2018bbd,
Bianchi:2018kzy, Bena:2018mpb, Bombini:2019vnc, Tian:2019ash,
Bena:2019azk, Bena:2019wcn}, some of which are reviewed in
section~\ref{sec:developments}.

For more explicit examples of single-mode superstrata -- original,
supercharged, and hybrid, see \cite{ Bena:2017xbt, Bena:2018mpb,
Bakhshaei:2018vux, Ceplak:2018pws, Heidmann:2019zws, Walker:2019ntz}.


\subsubsection{Asymptotically flat solution and conserved charges}

In order to extend the above asymptotically AdS superstrata to
asymptotically flat solutions that represent microstates of the D1-D5-P
black hole in flat space, we must add ``1'' to $Z_{1,2}$ as
\begin{align}
 Z_1\to 1+Z_1,\qquad
 Z_2\to 1+Z_2.
\end{align}
This does not affect the first-layer equations, but makes the
second-layer equations more complicated by introducing new source
terms.  
As a result, unlike asymptotically AdS ones, in asymptotically flat
superstrata, high-frequency sources in the second layer are not
completely canceled but are combined so as to remove singularities in
the solutions. This means that the coiffuring relations such as
\eqref{coiff_single-mode_hyb} get modified.
The asymptotically flat, single-mode solution with general $(k,m,n)$ was
found in \cite{Bena:2017xbt} for the original superstratum (except for
$\zeta$ in the RMS part), although the one for the hybrid superstratum
has not been written down as of writing.  For the original superstratum,
$\bh_4=0$, the coiffuring relation \eqref{coiff_single-mode_hyb} is just
\begin{align}
 b_1=b_4^2.\label{jczv7Jan20}
\end{align}
In the asymptotically flat version, this is modified to
\cite{Bena:2017xbt}
\begin{align}
 b_1={b_4^2\over 1+{a^2\over Q_5}{m+n\over k}}.
\end{align}
 In
the decoupling limit, ${a^2}\ll Q_{1,5}$, this relation falls back
to the AdS relation \eqref{jczv7Jan20}.

To read off asymptotic charges, we do not have to know the explicit form
of the asymptotically flat solution.  This is because the extra source
terms that appear in the asymptotically flat solution have a non-vanishing
wave number in the $v$ direction and thus vanish when integrated over
the $S^1$.  Therefore, we can read off charges from the ansatz quantities of
the asymptotically AdS solution \cite{Bena:2015bea}.
The D1- and D5-brane charges are simply $Q_1$ and $Q_5$, which are
related to the quantized numbers $N_1$ and $N_5$ as in
\eqref{Q1Q5_n1n5}.  The momentum charge $Q_p$ can be read off from 
\begin{align}
 {\cF} \sim -{2Q_p\over r^2}
\end{align}
and the angular momenta $\cJ_L,\cJ_R$ are read off from
\begin{align}
 \beta_{\phi}+\beta_{\psi}
 +\omega_{\phi}+\omega_{\psi}
 \sim \sqrt{2}\,{\cJ_L-\cJ_R\cos2\theta\over r^2}.
\end{align}
If we apply these relations to the single-mode hybrid superstratum, we
find
\begin{align}
 Q_p={m+n\over k}(B_4^2+\Bh_4^2),\qquad
 \cJ_L=R_y\Bigl[{a^2\over 2}+{m\over k}(B_4^2+\Bh_4^2)\Bigr],\qquad
 \cJ_R={R_y\over 2}a^2.
\label{gopv11Jan20}
\end{align}

The supergravity quantities $Q_p$ are related to the
quantized momentum number $N_P$ by
\begin{align}
 Q_p
 ={g_s^2\ap^4\over R_y^2 v_4}N_P
 ={Q_1 Q_5\over R_y^2 N}N_P,
\end{align}
where in the second equality we used
\eqref{Q1Q5_n1n5}.
On the other hand,  $\cJ_L,\cJ_R$ are related to the quantized angular momenta
$J_L,J_R$ by
\begin{align}
 \cJ_{L,R}
 ={g_s^2\ap^4\over R_y v_4}J_{L,R}
 ={Q_1 Q_5\over R_y N}J_{L,R}.
\end{align}

As mentioned before, the supergravity amplitudes $a,B_4,\Bh_4$ are
related to the CFT occupation numbers $N_0,N_{k,m,n},\Nh_{k,m,n}$
characterizing the state \eqref{gjpo11Jan20}.  Let us identify the
supergravity and CFT quantities as
\begin{align}
 a^2={Q_1 Q_5\over R_y^2}{N_0\over N},\qquad
 B_4^2={Q_1 Q_5\over R_y^2}{k N_{k,m,n}\over N},\qquad
 \Bh_4^2={Q_1 Q_5\over R_y^2}{k \Nh_{k,m,n}\over N}.
\label{jkap11Jan20}
\end{align}
Then the regularity constraint \eqref{gjtg11Jan20} becomes the strand
budget equation \eqref{irns7Jan20}, while \eqref{gopv11Jan20} translate
into
\begin{align}
 N_P=(m+n)(N_{k,m,n}+\Nh_{k,m,n}),\qquad
 J_L={N_0\over 2}+m(N_{k,m,n}+\Nh_{k,m,n}),\quad
 J_R={N_0\over 2},
\end{align}
which are exactly equal to the charges of the CFT state
\eqref{gjpo11Jan20}, giving a strong support for the holographic
dictionary.

\subsubsection{Multi-mode superstrata}
\label{sss:multi-mode_superstrata}

In the above, we focused on the single-mode hybrid superstratum, for
which coiffuring can be carried out explicitly. Namely, the
high-frequency source can be coiffured away by the
choice~\eqref{coiff_single-mode_hyb}, while the low-frequency source is just
the RMS mode that does not need coiffuring.  All the ansatz quantities
can be explicitly found for specific values of $\bm{k}=(k,m,n)$.

For the multi-mode superstratum \eqref{circular_class_1}, for the pair
$(\bm{k}_1,\bm{k}_2)$, there are four possible modes in the second-layer
source: the high-frequency mode $\bm{k}_1+\bm{k}_2$, the low-frequency
mode $\bm{k}_1-\bm{k}_2$, and the RMS mode
($\bm{k}_1-\bm{k}_1=\bm{k}_2-\bm{k}_2=\bm{0}$).  The RMS mode does not
need coiffuring.  The high-frequency mode can generally be coiffured away
\cite{Heidmann:2019zws}, as long as the supercharged modes are turned on.
On the other hand, how to coiffure the low-frequency mode is known only
on a case-by-case basis; see \cite{Bena:2015bea, Heidmann:2019xrd} for
explicitly worked out examples.

For asymptotic charges, only the RMS modes will contribute and the result
will be a simple sum over all modes; for example,
\eqref{gopv11Jan20} will be generalized to
\begin{align}
 Q_p=\sum_{k,m,n} {m+n\over k}\left[(B^{k,m,n}_4)^2+(\Bh^{k,m,n}_4)^2\right],
\end{align}
which is consistent with the CFT side with identifications
similar to \eqref{jkap11Jan20}.
Other charges will similarly be given by a sum over all modes.

\subsubsection{Superstrata based on other species}
\label{sss:strata_other_species}

\noindent
{\it Superstrata based on $\ket{++'}$}
\label{sss:strata_++'}

\noindent
In the above, we discussed superstrata based on the state $\ket{00}$.
At the linear level, the ``density mode'' $\ket{++'}$
in~\eqref{lin_spec_fixed_base} is another state that does not change the
base $\bbR^4$, and the superstratum based on it is also expected to have
flat base.

In \cite{Bena:2016agb}, a certain family of superstrata based on
$\ket{++'}$ were explicitly constructed. If one turns on density
fluctuation, generically, infinitely many modes will be turned on in the
harmonic functions $Z_{1,2}$.  However, if one turns on the $Z_4$ mode
at the same time in a coordinated way, the harmonic functions will
involve essentially only one mode.  This is called ``Style 1''
coiffuring in \cite{Bena:2016agb} (the superstratum based on $\ket{00}$ is called
``Style 2'' there).  More precisely, in their solution,
$(J^+_{-1})^k\ket{++}_{k+1}$, $(J^+_{-1})^k\ket{--}_{k-1}$ and
$(J^+_{-1})^k\ket{00}_k$ are turned on.\footnote{Their solutions include
generalization to excitations around the orbifold $({\rm AdS}_3\times
S^3)/\bbZ_p$ with $p\ge 1$, but here we are setting
$p=1$.}$^,$\footnote{Having a single mode turned on in the bulk means
that, on the boundary, infinitely many modes are turned on.  Namely, the
corresponding CFT state has $(J^+_{-1})^k\ket{++}_{k+1}$,
$(J^+_{-1})^k\ket{--}_{k-1}$ and $(J^+_{-1})^k\ket{00}_k$ turned on not
just for one value of $k$ but for all integer multiples of $k$.}

In \cite{Bena:2016agb}, only bosonic excitations on top of $\ket{++'}$
were considered but, as we can see from~\eqref{lin_spec_fixed_base},
their supercharged version based on $GG\ket{++'}$ are also expected to
have flat $\bbR^4$ base.

\medskip\noindent
{\it Superstrata based on $\ket{\dot{A}\dot{B}}$}

\noindent
In this article, we are restricting ourselves to superstrata that
preserve the symmetry of the internal manifold~$\cM$, for which the
supergravity fields take the form of \eqref{ansatzSummary}.  By relaxing
this condition and turning on fields that have legs along~$\cM$ (but are still independent of the coordinates of $\cM$), one can
construct superstrata \cite{Bakhshaei:2018vux} that are based on the
species $\ket{\dot{A}\dot{B}}$ listed in \eqref{T4_1-part_ch_pr}.  Such
solutions will include more scalars $Z_{I\ge 5}$ and forms $\Theta_{I\ge
5}$, corresponding to more tensor multiples of $d=6,\cN=(1,0)$
supergravity.

\subsection{Superstrata on 
the orbifold $({\rm AdS}_3\times S^3)/\bbZ_p$}
\label{ss:strata_on_AdS3xS3/Zp}

The superstrata reviewed above describe fluctuation around ${\rm
AdS}_3\times S^3$ (see section \ref{sss:empty_AdS3xS3}) which corresponds to
the circular profile \eqref{profile_emptyAdS3xS3} and whose CFT dual is
$[\ket{++}_1]^N$ in the R sector.
Instead, if we consider a $p$ times wound circle,
\begin{align}
 g_1+ig_2=a e^{2\pi i p\lambda/L},\qquad g_3+ig_4=g_5=0,\label{profile_emptyAdS3xS3/Zp}
 \qquad p\ge 1,
\end{align}
then we obtain the orbifold $({\rm AdS}_3\times S^3)/\bbZ_p$ whose
CFT dual in the R sector is
\begin{align}
  [\ket{++}_p]^{N/p}.\label{bfcg23Jan20}
\end{align}
In this case, the ansatz
data are somewhat changed from the ones given in
\eqref{ansatz_data_empty_AdS3xS3} to
\begin{subequations}
 \label{ansatz_data_empty_AdS3xS3/Zp}
 \begin{align}
 Z_1&={(p R_y)^2a^2\over Q_5\Sigma},\qquad
 Z_2={Q_5\over \Sigma},\qquad
 Z_4=0,\qquad \Theta_I=0,
 \\
 \beta&=  \frac{p R_y a^2}{\sqrt{2}\,\Sigma}(\sin^2\theta\, d\phi - \cos^2\theta\,d\psi)= p\beta_0,
 \\
 \omega&= \frac{p R_y a^2}{\sqrt{2}\,\Sigma}(\sin^2\theta\, d\phi + \cos^2\theta\,d\psi)=p \omega_0,\qquad \cF=0,
 \end{align}
\end{subequations}
and the relation
\eqref{aQR_y} is changed to
\begin{align}
 a^2={Q_1 Q_5\over (p R_y)^2}.\label{aQR_y/Zp}
\end{align}
We see that we can obtain the result for the $p$-wound case from the
original ($p=1$) case by the
replacement
\begin{align}
 R_y\to p R_y.
\end{align}
Because $R_y$ is the radius of identification for the $y$ circle, this
replacement means that we have a conical defect where the $y$ circle
shrinks.  Therefore the geometry is $({\rm AdS}_3\times S^3)/\bbZ_p$.
If we want to see the structure of the space more explicitly, we can do
the following coordinate transformation (cf.~\eqref{bulk_spectr_flow})
\begin{align}
 \phit
 =\phi-{t\over p R_y}
 ,\qquad
 \psit
 &=\psi-{y\over p R_y}
 \label{bulk_spectr_flow/Zp}.
\end{align}
The 6D metric becomes
\begin{align}
 ds^2_{6}
 &=\sqrt{Q_1Q_5}
 \left(-{r^2+a^2\over a^2 (p R_y)^2}dt^2+{r^2\over a^2 (p R_y)^2}dy^2
 +{dr^2\over r^2+a^2}
 +d\theta^2+\sin^2\theta\, d\phit^2+\cos^2\theta\, d\psit^2\right).
\label{fhug15Jan20}
\end{align}
Although this is locally ${\rm AdS}_3\times S^3$, because of
identification $(y,\phit,\psit) \cong(y+{2\pi R_y\over
p},\phit,\psit-{2\pi\over p})$, there is a $\bbZ_p$ singularity at
$r=0$.  This conical defect singularity is due to $p$ KK monopoles
sitting on top each other and is allowed in string theory.

We can say that the orbifolding is done by starting from a ``parent''
${\rm AdS}_3\times S^3$ and then quotienting the $y$-circle by $\bbZ_p$,
by making the radius $p$ times smaller.  This means that, if we consider
a fluctuation of the parent ${\rm AdS}_3\times S^3$ (namely, a
superstratum) and then divide the $y$-circle by $\bbZ_p$, then we get a
superstratum on $({\rm AdS}_3\times S^3)/\bbZ_p$, still with a
conical defect.  For this quotienting to make sense, the parent
superstratum must not have general $y$ Fourier mode numbers but only
ones that are single-valued after quotienting.  If we satisfy this
constraint, the resulting superstratum is expected to represent a valid
configuration in string theory.

This operation can be stated in the following way.  We consider the 
following CFT state
\begin{align}
\begin{split}
 \prod_{\psi,k,m,n} 
 \Bigl[(J^+_{-1})^m (L_{-1}-J^{3}_{-1})^n \ket{\psi}_k\Bigr]^{N^\psi_{k,m,n}}
 &=
 \prod_{\psi,k,m,n} 
 \Bigl[\ket{\psi;k,m,n}\Bigr]^{N^\psi_{k,m,n}}
 ~,\\
 \text{with}\qquad
 \sum_{\psi,k,m,n} k N^\psi_{k,m,n}&={N\over p}
\end{split}
\label{iuin14Jan20}
\end{align}
which, in the bulk, can be interpreted as a superstratum on the
parent ${\rm AdS}_3\times S^3$.  Note that the total strand length is
$N\over p$ and not $N$. It is assumed that $N$ is divisible by $p$,
namely, ${N\over p}\in\bbZ$.  Also, we require that $N^\psi_{k,m,n}=0$ unless ${m+n\over
p}\in\bbZ$.  For simplicity, we did not include states excited by
$G_{-1}^{+A}$, but including them is straightforward.
Now, given such a state, we define a state
\begin{align}
\begin{split}
 \prod_{\psi,k,m,n} 
 \Bigl[(J^+_{-{1\over p}})^m (L_{-{1\over p}}-J^{3}_{-{1\over p}})^n \ket{\psi}_{pk}\Bigr]^{N^\psi_{k,m,n}}
 &\equiv
 \prod_{\psi,k,m,n} 
 \Bigl[\ket{\psi;k,m,n;p}\Bigr]^{N^\psi_{k,m,n}}
 ~
 ,\\
 \text{with}\qquad
 \sum_{\psi,k,m,n} pk N^\psi_{k,m,n}&=N,
\end{split}
 \label{iukz14Jan20}
\end{align}
which is interpreted of as a superstratum on $({\rm AdS}_3\times
S^3)/\bbZ_p$.  Although involving fractional modes, this is a physically
allowed state of the orbifold CFT because of the restriction ${m+n\over
p}\in\bbZ$.  The bulk superstratum depends on $y$ as $e^{i(m+n)y/(p
R_y)}$ which is single-valued because of the restriction.

In \cite{Bena:2016agb}, a special family of single-mode superstrata of
this kind was explicitly constructed.  The dual CFT state was identified
to be the following:\footnote{They also present orbifolded superstrata
of ``Style 1''  mentioned in
section \ref{sss:strata_++'}.}
\begin{align}
 \Bigl[\ket{++}_k\Bigr]^{N_0}
 \Bigl[(J^+_{-{1\over p}})^{pk}\ket{00}_{p^2k}\Bigr]^{N_1}
 =
 \Bigl[\ket{++}_k\Bigr]^{N_0}
 \Bigl[\ket{00;pk,pk,0;p}\Bigr]^{N_1}
\end{align}
which is indeed of the form of \eqref{iukz14Jan20}. The state appearing
in the second factor has $\Delta L_0={1\over p}\cdot kp=k\in\bbZ$ and is
thus an allowed state of the orbifold CFT\@.  The explicit solution has
exactly the same form as the ordinary superstratum in
section~\ref{ss:explicit_strata} with $(k,m,n)\to (pk,pk,0)$ and with
$R_y\to p R_y$.  The geometry has a $\bbZ_p$ orbifold singularity just
as the $({\rm AdS}_3\times S^3)/\bbZ_p$ space which the solution is
excitation of.

This procedure expands the class of CFT states representable by
microstate geometries by inclusion of certain fractional modes.  These
states exist everywhere in the moduli space of the D1-D5 CFT, because
$({\rm AdS}_3\times S^3)/\bbZ_p$ contains only one non-trivial 3-cycle
\cite{Bossard:2019ajg}.

\section{Further developments}
\label{sec:developments}

Superstrata have provided a rich paradigm in which to study physics and
mathematics of black-hole microstructure.  Some of the physical aspects
of superstrata that have been explored are: explicit construction of
more general class of states, precision holography, relation to other
duality frames, scaling limits and asymptotically AdS$_2$ solutions,
scattering off superstrata, counting, and so on.  On the mathematics
side, some of the aspects that have been investigated include: the
structure of the BPS equations, integrability of the geometries,
structure of the ambi-polar base, etc.  Here, we will give a survey of
recent developments concerning superstrata and related subjects.

\subsubsection*{Generalizations and other duality frames}

The MSW black hole \cite{Maldacena:1997de} in five dimensions obtained
by compactifying M-theory on six-dimensional manifold and wrapping
M5-branes on 2-cycles in it is another prototypical black hole with
which to study black-hole microphysics.  The near-horizon geometry is
${\rm AdS}_3\times S^2$ and the dual CFT$_2$ is called the MSW CFT\@.
When the compactification manifold is $T^6$, a chain of duality
transformations relates the five-dimensional MSW system to the
six-dimensional D1-D5-P system.  The duality transformations involve
$T$-duality which requires an isometry direction along which the dual is
taken.  By this chain of duality,  D1-D5 superstrata with the
requited isometry direction can be mapped into superstrata in the MSW
system \cite{Bena:2017geu, Walker:2019ntz}, the latter being described
in five-dimensional supergravity.  This means that there is a map
between a subsector of the D1-D5 CFT and a subsector of the MSW CFT\@.
Empty ${\rm AdS}_3\times S^3$ in the D1-D5 system is mapped into a
D6-$\overline{\rm D6}$ configuration in the MSW system.  In the latter,
one can consider a gas of D0-branes \cite{Denef:2007yt,
deBoer:2008zn}\footnote{It was argued that these D0-branes puff out into
M2-branes whose Landau level degeneracy accounts for the entropy of the
MSW black hole \cite{Denef:2007yt, Gimon:2007mha}. These M2-branes are
supposed to wrap a non-trivial $S^2$ in the geometry and are sometimes
dubbed supereggs.  However, it was shown that such 
M2-branes will violate charge conservation~\cite{Martinec:2015pfa}
and/or break the supersymmetry~\cite{Tyukov:2016cbz} preserved by the
MSW black hole.  Therefore, these superegg M2-branes and their Landau
levels cannot be the precise description of the microstates.} which can
alternatively be represented by a gas of supergravitons in ${\rm
AdS}_3\times S^2$ \cite{deBoer:2009un}.  The MSW superstrata can be
regarded as  coherent states of this supergraviton gas.  Such
microstates in the MSW system are expected to be useful in understanding
the MSW CFT which remains mysterious.

As mentioned in \ref{sss:multi-mode_superstrata}, it is not yet known
how to construct general multi-mode superstrata including all
$(k,m,n,f)$.  In \cite{Heidmann:2019xrd}, interesting progress was made by
considering holomorphic superposition of some family of modes.  For
example, in section~\ref{sss:examples}, we discussed $(1,0,n)$
superstrata.  Let us consider superposing different modes in the
first-layer fields $Z_4,\Theta_4$.  For example,
\begin{align}
 Z_4=\sum_n b_4^{1,0,n}z_{1,0,n}
 ={R_y \over 2\Sigma}\left( \chi \sum_n  b_4^{1,0,n} \xi^n +{\rm c.c.}\right).
\end{align}
Here, we defined
\begin{align}
 \chi={a \over \sqrt{r^2+a^2}} \sin\theta\, e^{i\phi},\qquad
 \xi={r \over \sqrt{r^2+a^2}}e^{i{\sqrt{2}v\over R_y}},\qquad
 \eta={a\over\sqrt{r^2+a^2}}\cos\theta\, e^{i({\sqrt{2}v\over R_y}-\psi)}
\end{align}
in terms of which $\Delta_{k,m,n}e^{iv_{k,m,n}}=\chi^{k-m}\eta^m \xi^n$.
This suggests that we work with the holomorphic function
\begin{align}
 F(\xi)=\sum_n  b_4^{1,0,n} \xi^n,
\end{align}
rather than with the mode coefficients $b_4^{1,0,n}$. It turns out that
one can find the explicit expression for all fields in terms of
$F(\xi)$.  Moreover, regularity and no-CTC analyses can be completed in
terms of $F(\xi)$, and conserved charges can be written in terms of
$F(\xi)$.  In this $(1,0,n)$ case, there is no coiffuring needed but, in
other examples, such as the case (namely, holomorphic superposition of
the $(k,0,1)$ superstratum with general coefficients $b_4^{k,0,1}$),
both high- and low-frequency coiffuring can be done in terms of
holomorphic functions.  Having arbitrary holomorphic functions allows us
to construct physically interesting solutions, such as ones with a
momentum wave localized in the $y$-circle direction, unlike the
single-mode superstratum in which the momentum wave is delocalized in
the $y$ direction.  Turning on all possible $(k,m,n)$ modes corresponds
to having holomorphic functions that depend on all three variables,
$\chi, \xi, \eta$.  Explicitly constructing the solution would fulfill
the promise of general superstrata being parametrized by three
variables.  Achieving that goal may be quite difficult technically, but
having holomorphic functions of one variable is already a promising step
forward.

It is interesting to see if the superstratum technology can be
generalized to a non-super\-symmetric setting. At the linear level, this
is straightforward with AdS$_3$ asymptotics, because all one must do is
to use the solution-generating technique to act on a chiral primary
state not only with left-moving generators $L_{-1},G^{+A}_{-1},J^+_{-1}$
but also with right-moving ones $\Lt_{-1},\Gt^{+A}_{-1},\Jt^+_{-1}$.  In
\cite{Roy:2016zzv, Bombini:2017got}, this was carried out and certain
non-supersymmetric microstate geometries were constructed.  Furthermore,
those geometries were extended to asymptotically-flat solutions by a
matching technique.  It is desirable to extend such non-supersymmetric
superstrata to non-linear solutions; ideas used to construct
five-dimensional non-supersymmetric microstates~\cite{Bena:2009fi} may
be useful for such extension.  Just as in the supersymmetric fluctuation
modes discussed in section~\ref{sss:linear_spectrum}, some particular
non-supersymmetric modes are expected to be simpler and 
technically easier to construct
than others
\cite[Appendix C]{Ceplak:2018pws}.

By $S$-duality, the D1-D5 system is related to the F1-NS5 system, which
upon $T$-duality becomes the P-NS5 system.  In these duality frames,
only NS-NS fields are turned on and an exact worldsheet CFT description
of the background is possible.  In \cite{Martinec:2017ztd,
Martinec:2018nco, Martinec:2019wzw}, the relevant worldsheet CFT was
constructed, and the spectrum of the fundamental string and possible
D-branes were analyzed. This duality frame may be useful in studying the
bulk realization of the fractional and higher modes that are crucial in
understanding the microstates of the D1-D5-P (or the F1-NS5-P) black
hole.

\subsubsection*{Precision holography}

Studying the correlation function in a microstate and matching it
between bulk geometries and CFT states is sometimes called ``precision
holography''.  Precision holography was developed for 1/4-BPS
microstates (Lunin-Mathur geometries) in \cite{Kanitscheider:2007wq,
Kanitscheider:2006zf, Skenderis:2008qn} and extended to 1/8-BPS
microstates (superstrata) in \cite{Giusto:2015dfa, Giusto:2019qig}.
According to the non-renormalization theorem in \cite{Baggio:2012rr},
correlation functions of type $\ev{\cO_{1/8}\cO_{1/4}\cO_{1/8}}$ are not
renormalized, where $\cO_{1/8}$ and $\cO_{1/4}$ are 1/8-BPS and 1/4-BPS
operators.  If $\cO_{1/8}=H$ is a backreacted 1/8-BPS geometry (a heavy
state with dimension of order $N$) and $\cO_{1/4}=L$ is a light probe (a
chiral primary operator with dimension of order one), then
$\ev{HLH}=\ev{H|L|H}$ can be found by computing the one-point function
in the superstratum.
In \cite{Giusto:2015dfa}, the one-point function of dimension-one chiral
primary operators was studied and, in particular, the existence of the
term in $Z_1$ necessary for coiffuring low-frequency source was
confirmed from CFT\@.  The existence of this term was shown
\cite{Giusto:2015dfa} to be consistent also with holographic
entanglement entropy.
In \cite{Giusto:2019qig}, the one-point
function of dimension-two operators were studied and the existence of
the $\cO(b^2)$ term in $Z_1$ was shown to be due to mixing between
single-particle operators and two-particle operators in the holographic
dictionary.  These results are quite intriguing, because the relevant
terms in the supergravity ansatz originate from regularity, which
requires information about the interior of spacetime, whereas CFT
computation only involves operators of small dimension, which are
related to deformations of the geometry near the boundary.  This
demonstrates the power of CFT in predicting non-trivial features of the
dual geometry.

Computing the bulk two-point function of a light probe operator in a
heavy backreacted geometry gives correlation function of type
$\ev{H|LL|H}=\ev{HLLH}$.\footnote{This correlation function, being really a
four-point function, is not protected.}  Two-point functions decay in a
black-hole background, because the bulk field gets absorbed into the
horizon.  It is interesting to see whether and how the correlation
function in microstate geometries mimics such behavior.  Although the
two-point function in a microstate geometry is expected to decay at
initial times, it should not go to zero and must show recurrence after a
long but finite time.
In \cite{Galliani:2017jlg, Bombini:2017sge}, the correlation function in
1/4-BPS geometries of the type given in section~\ref{ss:LM_Z4} was
studied.  The computation was mostly done in the ``shallow'' limit $b\ll
a$, although in \cite{Bombini:2017sge} an exact expression valid for any $b$ was obtained
for $k=1$.  
For 1/8-BPS microstates, {\it i.e.\/}~superstrata, two-point function
was  studied in \cite{Raju:2018xue, Bombini:2019vnc, Tian:2019ash}.
Most of these works focused on the shallow limit but, in
\cite{Bena:2019azk}, the deep throat limit of the $(1,0,n)$ superstratum
was studied using a WKB technique, and it was found that two-point
functions decay as in the BTZ black hole for
$t\lesssim\sqrt{N_1N_5}R_y$, while for $t\sim N_1N_5 R_y$ large echoes
are coming back from the cap.  Because this is an atypical state, the
echo is strong and the correlation function comes back to almost the
original value.  However, more general superstrata are expected to show
less spiky behavior.

Other work on the holography of 1/8-BPS microstates includes
\cite{Giusto:2018ovt, Tormo:2019yus}.

\subsubsection*{Further   aspects of superstrata}

Some supersymmetric microstate geometries have been argued to have
non-linear instability~\cite{Eperon:2016cdd, Keir:2016azt,
Eperon:2017bwq}, which suggests that they want to evolve into more
typical microstates \cite{Marolf:2016nwu}.  In~\cite{Raju:2018xue},
based on statistical-mechanical considerations, it was argued that
microstate geometries that are classically distinguishable from the
black-hole geometry are atypical. Superstrata on ${\rm AdS}_3\times S^3$
are not typical microstates of the three-charge black hole, because
their entropy is parametrically smaller than the black-hole entropy
\cite{Shigemori:2019orj}.  

However, these do \emph{not} mean that those
superstrata are irrelevant in studying the microscopic physics of black
holes. 

First, by studying their instability, superstrata must give us
information about the nature of more typical microstates that they have
tendency to evolve into. 
For the BTZ black hole, the tidal force felt by an object falling into
it can be made small for a large black hole, even at the horizon.
However, for a capped BTZ geometry such as superstrata with a deep
throat, it was found that, generally, an infalling object experiences a
Planckian tidal force \cite{Tyukov:2017uig, Bena:2018mpb}.  The tidal
force is $|\cA|_{\rm throat}\sim a^2Q_P^2/(\sqrt{Q_1 Q_5\,}\,r^6)$,
where $a$ is the scale of the cap determined by the residual angular
momentum, which vanishes in the BTZ limit, $a\to 0$.  This large stress
force comes from the deviation of the microstate geometry from BTZ
geometry, amplified by the relativistic speed of the infalling particle.
This means that a particle dropped into a superstratum with a deep
throat gets stretched into a string and/or ripped into strings.  By
following the fate of the string(s), we must be able to get a hint as to
the more typical microstates that the superstratum wants to evolve into.
If the tidal force tend to transform the particle into a massive string by
exciting string oscillator modes on it, that would mean that
supergravity is not enough for describing microstates and stringy modes
must be included.  Instead, if the tidal force turns the particle into
many massless strings, that would mean that supergravity is still good
but we must consider more general geometries than the known superstrata.

Secondly, even if atypical, they do behave just as a black hole to
certain probes; by studying response to such probes we can learn how the
non-unitary behavior of black holes emerges from unitary behavior of
microstates.  For research in this direction see \cite{Galliani:2017jlg,
Bombini:2017sge, Raju:2018xue, Bombini:2019vnc, Tian:2019ash,
Bena:2019azk} mentioned above.

Although superstrata naturally come with AdS$_3$ asymptotics, one can
take a scaling limit of deep superstrata geometries, such as the
$(1,0,n)$ stratum, and find an asymptotically-AdS$_2$ superstratum
\cite{Bena:2018bbd}.  Black-hole microstates with AdS$_2$ asymptotics
are interesting, particularly because of the claim
\cite{Dabholkar:2010rm, Chowdhury:2015gbk} that black-hole microstates
must have zero angular momentum in four dimensions ($J_R=0$ in five
dimensions) and fit in an AdS$_2$ region, and also because of the recent
surge of interest in the near-AdS$_2$/SYK correspondence 
\cite{Sachdev:1992fk, Kitaev} 
(for a review
of the already large literature see {\it e.g.\/}~\cite{Sarosi:2017ykf}).  One cannot
have excitation in global AdS$_2$~\cite{Maldacena:1998uz,
Almheiri:2014cka} because a finite excitation makes the dilaton diverge
at the end of the space.  However, the existence of a capped AdS$_2$
does not contradict with the no-go theorem, because the divergent
dilaton is interpreted as the collapsing of an internal $S_1$
which caps off the geometry. In~\cite{Bena:2018bbd}, it was found that
the non-supersymmetric excitations at the bottom of the capped AdS$_2$
superstratum are normalizable with spectrum $\Delta E=4J_R/(N
R_y)$,\footnote{ The spectrum in the $(1,0,n)$ geometry was studied in
\cite{Raju:2018xue} before.}  which for $J_R=1/2$ reproduces the CFT
expectation.  However, it remains to be seen if these excitations
preserve the AdS$_2$ asymptotics, when backreacted.  
For examples of five-dimensional microstate
geometries with AdS$_2$ asymptotics and their relevance to 
pure-Higgs branch states 
\cite{Bena:2012hf}
of the dual quiver quantum mechanics,
 see
\cite{Heidmann:2018vky}.

Further aspects of superstrata and microstate geometries studied in the
literature include: scattering off microstate geometries
\cite{Bianchi:2017sds, Bianchi:2018kzy}; trailing string and drag force
\cite{Gubser:2006bz} in microstate geometries \cite{Bena:2019wcn};
integrability of the superstratum backgrounds \cite{Bena:2017upb,
Walker:2019ntz}; ambi-polar hyper-K\"ahler space, pseudo-harmonic from,
and prepotentials of five-dimensional microstate geometries
\cite{Bena:2017geu, Walker:2019ntz, Tyukov:2018ypq}.

\section{Concluding remarks}
\label{sec:conclusion}

In this article, we reviewed aspects of superstrata, a large family of
microstate geometries of the D1-D5-P black hole.  They can be
constructed systematically using the linear structure of BPS equations
and represent coherent states of 1/8-BPS supergravitons. They provide an
ideal setup in which to study the physics of black holes.  Although
their holographic dictionary is well understood and has a deceivingly
simple structure, their bulk physics is surprisingly  rich; for example, some
superstrata have a long throat with a large redshift and a small gap,
which is quite non-trivial from the CFT viewpoint.

As already mentioned, superstrata on ${\rm AdS}_3\times S^3$ involve, in
CFT language, only rigid-generator descendants of chiral primary states,
and their entropy is not enough to account for the entropy of the
D1-D5-P black hole \cite{Shigemori:2019orj}. To reproduce the full
entropy, it is crucial to understand the bulk realization of fractional
and higher modes mentioned in section~\ref{ss:general_1/8-BPS_states}.
Some of fractional modes are realized in superstrata on the orbifold
$({\rm AdS}_3\times S^3)/\bbZ_p$ as reviewed in
section~\ref{ss:strata_on_AdS3xS3/Zp}.  However, these represent only a
special kind of fractional mode on a limited class of chiral primary
states; understanding of general fractional modes on general chiral
primary states is still missing.  Also, the bulk realization of general
higher modes is poorly understood (see however \cite{Lunin:2012gp,
Giusto:2013bda}).  Although multi-center superstrata may correspond to
such states \cite{Bena:2014qxa}, it is also possible that intrinsically
stringy excitations are essential.  In any case, this is one of the most
important problems that have to be resolved in order to further our understanding
of black-hole microstates.

Superstrata have helped deepen our understanding of black holes by their
rich physical and mathematical content. Further investigation is bound
to reveal more surprising aspects of superstrata and lead to better
understanding of the microscopic working of black holes.

\section*{Acknowledgments}

\vspace{-2mm}

I thank Iosif Bena, Nejc \v{C}eplak, Stefano Giusto, Emil Martinec,
Rodolfo Russo, David Turton and Nick Warner for fruitful collaborations
and for sharing illuminating insights.
I also thank Pierre Heidmann, Daniel Mayerson and Alexander Tyukov for
valuable discussions.
This work was supported in part by JSPS KAKENHI Grant Numbers
16H03979, and MEXT KAKENHI Grant Numbers 17H06357 and 17H06359.

\appendix


\end{document}